\documentclass[aps,prb,twocolumn]{revtex4-2}
\usepackage{bm}
\usepackage{graphicx}
\usepackage{xcolor}
\usepackage{amsmath}
\usepackage{mathtools,amsfonts,amssymb,amsthm}
\colorlet{darkred}{red!85!black}
\colorlet{darkgreen}{green!50!black}
\colorlet{darkblue}{blue!60!black}
\usepackage{hyperref}
\hypersetup{
	colorlinks   = true, 
	urlcolor     = darkblue, 
	linkcolor    = darkgreen, 
	citecolor   = darkred 
}
\begin{document}
\title{Quantum and classical Shapiro steps in small Josephson junctions}

\author{Miriam Resch}
\author{Joachim Ankerhold}
\address{Institute for Complex Quantum Systems and IQST, Ulm University, Albert-Einstein-Allee 11, D-89069 Ulm, Germany}

\author{Brecht I. C. Donvil}
\affiliation{Hensoldt, W\"orthstra\ss e 85, 89077 Ulm}

\author{Paolo Muratore-Ginanneschi}
\address{Department of Mathematics and Statistics, University of Helsinki, P.O. Box 68, 00014 Helsinki, Finland}

\author{Dmitry Golubev}
\address{HQS Quantum Simulations GmbH, Rintheimer Str. 23, 76131 Karlsruhe, Germany}

\begin{abstract}
We propose a model describing the formation of both dual (quantum) and classical Shapiro steps in small
Josephson junctions. According to this model, the dual Shapiro steps are formed at relatively low frequency of the microwave signal and low microwave power, while the classical steps are formed in the opposite limit of
high frequency and power. The crossover between the two regimes is
controlled by a single parameter -- the effective relaxation time of the environment.
The model accounts for the effect of a large inductor in the bias
circuit, which has been used in recent experiments to protect the junction from the high frequency noise of the environment.
We predict the possibility of observing both types of steps in the same sample. 
Our model describes the I-V curves observed in the experiments with reasonable accuracy, thus
opening up an opportunity for quantitative fitting of the data.
\end{abstract}

\maketitle

\section{Introduction}
One of the main building blocks in superconducting circuits are Josephson junctions. They introduce the wanted nonlinearity to explore in combination with linear elements such as inductors, capacitances, and ohmic resistors  a plethora of fundamental quantum phenomena. In addition, they serve as central elements for quantum technological applications, for example, in quantum computing architectures.

The physics of Josephson junctions is governed by two conjugate degrees of freedom, namely, phase and charge. Their fluctuation properties are not only determined by Heisenberg's uncertainty relation but by the electromagnetic environment of the entire circuitry. In the domain, where the dynamics of the junction is ruled by the phase as a classical variable, 
Shapiro steps, discovered more than 60 year ago \cite{Shapiro}, appear in the current-voltage characteristics when it is
subject to a microwave radiation at dc-voltages  
\begin{eqnarray}
V_n = \frac{\hbar\omega_0}{2e}n,\ n\, \text{integer}\, .
\label{steps}
\end{eqnarray}
Here $\omega_0$ is the angular frequency of the radiation and $e$ is the electron charge.
Shapiro steps are caused by the nonlinear nature of the Josephson current-phase relation, $$I_J(\varphi)=I_C\sin\varphi,$$
which leads to synchronization between the microwave radiation and Josephson oscillations. 

It has been  predicted later \cite{LiZo1985,Averin,AL} that
in small area Josephson junctions, which are sufficiently well protected from the noise of the environment, another
type of Shapiro steps can be observed for dc-current biased junctions subject to an ac-current drive. These are the so-called dual Shapiro steps, or quantum Shapiro steps,
which appear at fixed current values
\begin{eqnarray}
I_n = 2e f_0 n,
\label{In}
\end{eqnarray} 
where $f_0 = \omega_0/2\pi$. These dual steps are caused by synchronization between the microwaves and the Bloch oscillations in the lowest Bloch band of the periodic Josephson potential with the phase as a quantum degree of freedom. Dual Shapiro steps have been first experimentally observed in Josephson junctions \cite{Kuzmin}. 
In this first experiment, the quantum steps were strongly smeared by noise and could be only seen in the differential resistance $dV/dI$. 
More recently, sharp dual steps, directly visible on the I-V curve, have been demonstrated in inductively protected phase slip superconducting nanowires \cite{PS1}, in single Josephson junctions \cite{PS2,PS3} and in single Cooper pair transistors \cite{PS4}.

The theory of the dual Shapiro steps has been developed over many years. 
It has been understood from the very beginning \cite{Averin} that quantum steps can only be observed if the effective resistance of the ohmic environment exceeds the
resistance quantum for Cooper pairs $$R_Q=\frac{h}{4e^2}=6.453\;  \mathrm{k}\Omega,$$ see e.g. Refs \cite{Hekking,Glazman} for more details. 
Subsequently, the effect of heating and quantum fluctuations on the widths of the dual steps has been analyzed \cite{GZ,Hekking,Glazman} and the
noise properties associated with Bloch oscillations have been studied \cite{GZ,Glazman}. 
The importance of inductance for protecting the junction from
high-frequency noise has been first realized by experimentalists,
and subsequently its effect has been investigated theoretically \cite{Hassler}. 
Most theory models rely on the tight binding approximation for the periodic Josephson potential, which leads to
cosine dispersion of the lowest Bloch band, and neglect Zener tunneling to the higher Bloch bands. 
The effect of Zener tunneling on the I-V curve of a small Josephson junction has been
considered, for example, in Refs. \cite{ZK,GZ1}, but only in the DC bias regime, where Shapiro steps are absent. 
Recent progress in experiment calls for the development of more realistic models suitable
for describing the experimental data in wide range of parameters. An important step towards this goal has been taken in
Ref. \cite{Vora}, where Monte Carlo simulations of the dual Shapiro steps have been carried out, 
and unwanted effects like single electron tunneling and Zener tunneling have been included. 
As a result, good agreement with the experiment \cite{Kuzmin} has been achieved.
Here we follow a similar route and develop a model of dual and classical Shapiro steps in
small Josephson junctions,
which includes Zener tunneling and, in addition, accounts for the effect of the large
inductance used in the experiments to protect the junction.
We ignore quasiparticle tunneling because in experiments the dissipation
predominantly comes from bias resistors.
We find that our model
describes recent experiments \cite{PS2,PS3} rather well, even though it overestimates the
size of the dual steps in the sample with large critical current.

The paper is organized as follows: in Sec. \ref{model} we make a brief introduction to the problem
and present our model, in Sec. \ref{results} we compare our results with the experiments \cite{PS2,PS3},
and in the Appendices we present the technical details.

\section{Model}
\label{model}

\begin{figure}
\includegraphics[width=0.9\columnwidth]{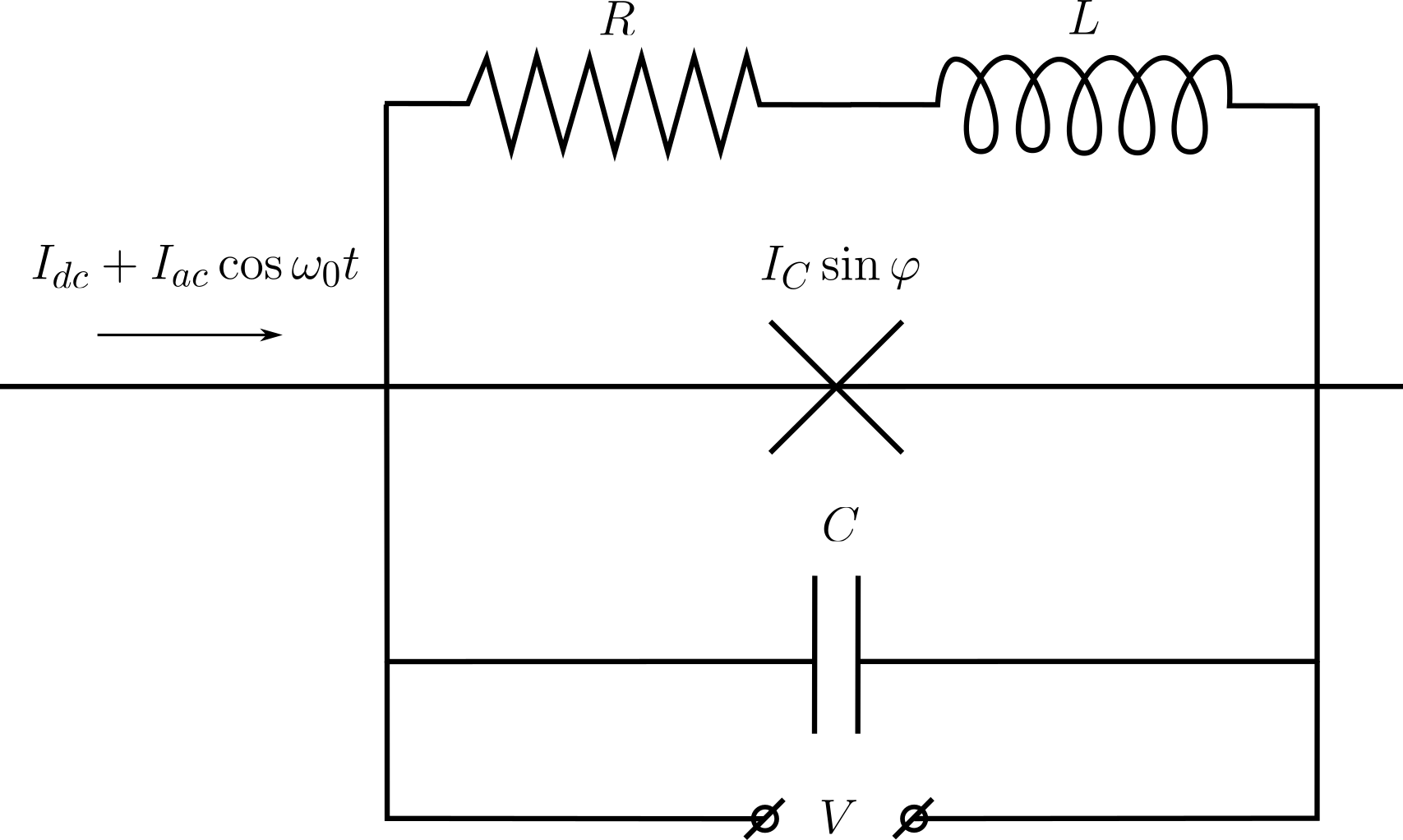}
\caption{Resistively and capacitively shunted Josephson junction (RCSJ model). The junction is current biased
with the current $I_{\rm dc} + I_{\rm ac}\cos\omega_0t$ and the voltage drop across the junction $V$ is measured by a voltmeter. }
\label{Fig1}
\end{figure}

We consider a small Josephson junction shunted by a capacitor and by a resistor connected in series with an inductor. The bias current containing both DC and AC components is applied from the current source and the voltage drop across the junction is measured. The corresponding circuit is depicted in Fig. \ref{Fig1}. An electrically equivalent circuit with the voltage source is
shown in Fig. \ref{Fig2}. The latter circuit has been used in the experiments \cite{PS1,PS2,PS3,PS4}.
In both circuits the inductor plays a crucial role of cutting off the high frequency noise, and it has been 
experimentally established that sharp dual Shapiro steps cannot be observed without it.

\subsection{Duality between classical and quantum Shapiro steps}

\begin{figure}
\includegraphics[width=0.9\columnwidth]{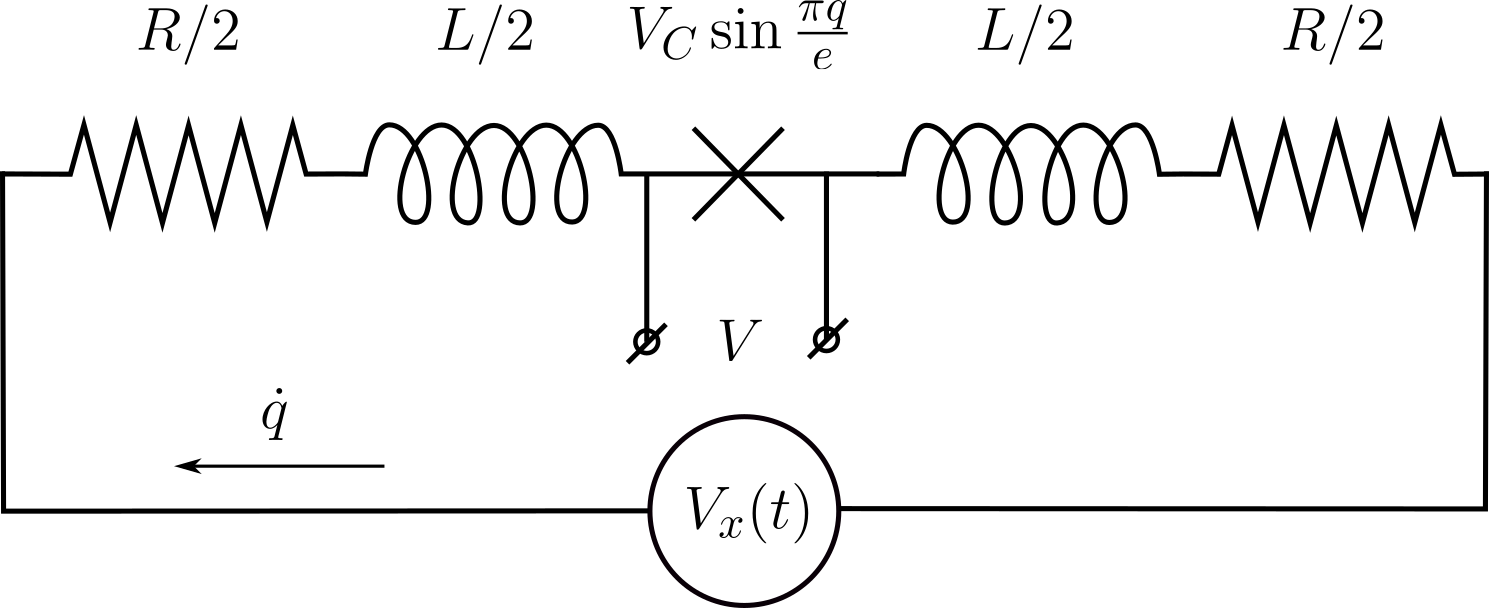}
\caption{Setup of the experiments \cite{PS1,PS2,PS3,PS4}, which is favorable for observation of the dual Shapiro steps. The junction is voltage biased
with the voltage $ V_x(t) = V_{\rm dc} + V_{\rm ac}\cos\omega_0t$ via the protecting resistors and inductors. 
The voltage drop across the junction $V$ is measured by a voltmeter and plotted versus the average current $\langle \dot q \rangle$. }
\label{Fig2}
\end{figure}

In this section we briefly explain the origin of dual Shapiro steps in simple terms.

We choose the circuit of Fig. \ref{Fig1} because it is easier to model. In a lumped element description it consists  of a capacitance $C$,  a shunt resistance $R$, an inductance $L$ connected in series with the resistor $R$, and a Josephson junction with critical current $I_C$. 
The classical Shapiro steps 
are then described by the equation of motion for the Josephson phase $\varphi$, i.e. 
\begin{equation}
\begin{split}
\frac{C \,\hbar}{2\,e} \ddot{\varphi}(t)&+ \int_{-\infty}^t dt'\, Y(t-t')
\frac{\hbar\,\dot{\varphi}(t')}{2e} + I_C\sin\varphi(t) 
\\
&=\, I(t) + \xi(t)\, ,
\end{split}
\label{RSJ}
\end{equation}
with the response function  
\begin{equation}
    Y(t)=\frac{e^{-R\, t/L}}{L}
\end{equation}
 whose Fourier transform is the admittance $Y(\omega)=1/Z(\omega)$ with impedance $Z(\omega)=R-i\omega L$.
The time-dependent bias current is given by
\begin{eqnarray}
I(t) = I_{\rm dc} + I_{\rm ac}\cos\omega_0t 
\label{bias}
\end{eqnarray}
and the noise process $\xi(t)$ is Gaussian  with zero mean value and correlation function
\begin{eqnarray}
\langle \xi(t)\xi(0) \rangle = k_B T\int \frac{d\omega}{\pi} e^{-i\omega\,t}  {\rm Re}\{Y(\omega)\}
\end{eqnarray}
with temperature $T$ of the resistor. This correlation function is related to the admittance $Y(\omega)$ via the fluctuation dissipation theorem.

In the quantum regime, the bare current-biased Josephson junction is described by the standard Hamiltonian
\begin{eqnarray}
H_J = - 4E_C {Q^2} + E_J(1-\cos\varphi) - \frac{\hbar I(t)}{2e}\varphi
\label{HJ}
\end{eqnarray}
with the charging energy
$$E_C = \frac{e^2}{2\,C}$$ 
and the Josephson energy
$$E_J=\frac{\hbar \,I_C}{2\,e}$$
as the relevant energy scales and the conjugate charge and phase operators obeying $[\varphi, Q]=i 2e$. 

In absence of a bias current $I(t)=0$, the Hamiltonian (\ref{HJ}) is widely used in applications, for instance, to model superconducting qubits see e.g. \cite{KrKjYaOrGu2019}. Due to its $2\pi$-periodicity in $\varphi$, the Bloch theorem applies and predicts the existence of energy bands that are $2e$-periodic with respect to the quasi-charge $q$ corresponding to $Q$. In the limit $E_J\gg E_C$ (deep energy wells of the $\cos$-potential) the lowest energy band is obtained as
\begin{eqnarray}
E_0(q) = \epsilon_0 - \delta_0\cos\frac{\pi q}{e}, 
\label{E0}
\end{eqnarray}
where
$$\epsilon_0 = \sqrt{2\,E_J\,E_C} - \frac{E_C}{4}$$ is the energy of the lowest level at the bottom of the cosine potential well, and
\begin{eqnarray}
\delta_0 = 16\sqrt{\frac{E_JE_C}{\pi}} \left(\frac{E_J}{2E_C}\right)^{1/4} \, e^{-\sqrt{8E_J/E_C}}
\label{delta0}
\end{eqnarray}
is the half-bandwidth corresponding to the tunneling between the wells. The next higher-lying energy band is separated from the lowest band by an energy on the order of $\sqrt{8\,E_J\,E_C}$.  The physical meaning of $q$ is that of the charge transferred through the junction \cite{Averin,AL,SZ}. For a given $q$ the average voltage drop across the junction follows
\begin{eqnarray}
V = \frac{dE_0}{dq} = V_C\sin\frac{\pi q}{e},
\label{VB}
\end{eqnarray}
where $V_C = \pi\delta_0/e$ is the critical voltage. 

Now, at non-zero bias current $I(t)\neq 0$, the quasicharge becomes time dependent and the dual Shapiro steps originate from the synchronization of Bloch oscillations in the lowest energy band (\ref{E0}) with an external microwave signal \cite{Averin,AL} at the current values (\ref{In}). In this situation the relevant degree of freedom is the quasi-charge which follows, for the circuit depicted in Fig. \ref{Fig2}, the equation of motion \cite{LiZo1985,Averin,AL}
\begin{equation}
\begin{split}
L\,\ddot{q}(t) &+ R \,\dot{q}(t) +  V_C\,\sin\frac{\pi q(t)}{e} 
\\
&= V_{\rm dc} + V_{\rm ac}\,\cos\omega_0 t + \xi_v(t),
\end{split}
\label{eq_ps}
\end{equation}
where  $\xi_v(t)$ is the voltage noise of the resistor. 
To obtain the I-V curve, one solves Eq. (\ref{eq_ps}) in order to compute the time and ensemble average (which here and below we denote by $\langle \cdot \rangle$) of the time derivative of the quasicharge, i.e. the DC current $I_{\rm dc}$ through the junction, from
$$I_{\rm dc}=\overline{\langle\, \dot {q} \,\rangle}=\lim_{t\nearrow\infty}\frac{1}{t}\int_{0}^{t}\mathrm{d}s \,\dot {q}(s).$$ 
Equation (\ref{eq_ps}) is valid if the system always stays in the lowest Bloch band, i.e.
if Zener tunneling to the upper bands can be neglected.

Apparently, Eqs. (\ref{eq_ps}) and (\ref{RSJ}) are mathematically very similar. This is the origin of the the well-known 
duality between the classical and quantum regimes in the dynamics of a Josephson junction. The similarity becomes even more pronounced
when we consider the high temperature (and $L\to 0$) limit of (\ref{RSJ}) whilst holding the ratio $k_B\,T/R$ fixed. Then one has for the kernel $Y(t)\propto \delta(t)$ so that the noise tends towards white noise and  (\ref{RSJ}) reduces to a time-local Langevin equation with linear friction as (\ref{eq_ps}). This high temperature limit is often referred to as the resistively and capacitively shunted junction (RCSJ) model. It accurately describes the phenomenon of underdamped phase diffusion in Josephson junction \cite{KiNiClBuHePe2005}.

 Shapiro steps then emerge according to the following simple arguments. For the standard case, one uses the second Josephson relation  between the total voltage drop across the junction and the phase velocity, namely,  
 $$V(t)=\frac{\hbar\,\dot{\varphi}(t)}{2\,e},$$ 
 assuming that $V$ is the sum of a dc-bias and an RF component of frequency $\omega_{0}$. Then, a simple heuristic argument \cite{ShJaHo1964} shows that classical steps (\ref{steps}) occur at the resonances $2 e V_{\rm dc}=n\hbar\omega_0$ which implies
$$\overline{\langle\dot{\varphi}\rangle} = \omega_{0} \,n.$$ 

For the dual case,  Eq. (\ref{eq_ps})  translates this condition into 
$$\pi\,\overline{\langle\, \dot {q}\,\rangle} = \omega_0\,n\,e,$$ 
whence the positions of the dual steps (\ref{In}) follow. We refer to \cite{LiZo1985,Averin,AL} for a first-principles derivation of these results. 
  Dual Shapiro steps are observed at current values below the critical
current $I_C$ or, more precisely, below the switching current $I_{\rm sw}<I_C$, i.e.\ $I_{\rm dc} < I_{\rm sw}$, and at small voltages, where the system mostly occupies the lowest Bloch band. In contrast, classical Shapiro steps occur at higher currents and voltages, where higher bands become occupied.

\subsection{Kinetic equation for charge distribution}

In this section we extend the above quantum mechanical model of the bare Josephson junction to include the impact of the impedance wihtin a system+reservoir model. It is assumed that $E_J$ is sufficiently smaller than $E_C$ which in turn allows to apply a perturbative treatment. For a pure dc-bias, this eventually leads to the known $P(E)$-theory where $P(E)$ captures the ability of a reservoir in thermal equilibrium to absorb or emit energy quanta with the junction. In presence of an RF-field, this setting must be generalized to an effective equation of motion for the marginal probability distribution of the quasicharge. 

Accordingly, we start with the total Hamiltonian of the circuit in Fig. \ref{Fig1} described by
\begin{eqnarray}
H = H_J + H_B + V,
\label{H}
\end{eqnarray}
with the Hamiltonian of the Josephson junction $H_J$ in Eq. (\ref{HJ}), the Hamiltonian of the bath of oscillators with coordinates $x_k$, momenta $p_k$, masses $m_k$ and frequencies $\omega_k$ describing the shunt resistance $R$ and the inductance $L$, respectively, i.e. 
\begin{eqnarray}
H_B = \sum_k\left( \frac{p_k^2}{2m_k} + \frac{m_k\omega_k^2 x_k^2}{2} \right)\, ,
\end{eqnarray}
and the interaction
\begin{eqnarray}
V = -\sum_k \left( c_k x_k\varphi - \frac{c_k^2}{2m_k\omega_k^2}\varphi^2 \right)\, .
\end{eqnarray}   
The coefficients $c_k$
define the quasi-continuous spectrum of bath oscillators via the spectral density $J(\omega)$ as 
\begin{eqnarray}
J(\omega) = \sum_k \frac{\pi\,c_k^2}{2\,m_k\,\omega_k}\delta(\omega-\omega_k) = \frac{\hbar^2\omega}{4e^2}\frac{R}{R^2+\omega^2L^2},
\end{eqnarray}
which is proportional to real part of the admittance $Y(\omega)$. 

With the total Hamiltonian at hand, the crucial step is now to consistently eliminate the bath degrees of freedom. This requires the path integral approach as outlined in \cite{Schon,SZ} in order to account for the quantum non-Markovian effects important at low temperature.
To make the problem tractable, we make a set of approximations and proceed as discussed in detail in App.~\ref{sec_Gamma}. In essence, the nonlinear  action of the reduced system (junction) is expanded in lowest order in $E_J$ and then, by taking time derivatives, the path integral is cast into a time-local evolution equation for the charge distribution $W(t,Q)$ where the charge $Q$ accumulated in the capacitor $C$ appears as the conjugate variable to the quasi-classical Josephson phase. Non-Markovian effects are taken into account in form of  renormalized parameters. 
This way, we arrive at
\begin{eqnarray}
&& \frac{\partial W}{\partial t}(t,Q)  = 
\frac{\partial}{\partial Q}\left( \frac{Q}{RC} - I^*(t)+ \frac{k_BT^*}{R} \frac{\partial}{\partial Q}\right)W(t,Q)
\nonumber\\ && 
+\, \Gamma(t,Q+2e)W(t,Q+2e) + \Gamma(t,-Q+2e)W(t,Q-2e)
\nonumber\\ &&
-\, \Big{(}\Gamma(t,Q) + \Gamma(t,-Q)\Big{)}W(t,Q).
\label{kin}
\end{eqnarray}
Before we explain in detail the basic ingredients of this equation, let us first comment on its general structure. Apparently, the first line describes quasi-classical overdamped charge diffusion of the RC-elements of the circuit (Smoluchowski equation) with effective bias current $I^*(t)$ and effective temperature $T^*$. The Cooper pair transfer across the junction is captured in the second and third line  by time-dependent golden rule type of rates. As we will discuss below, this hybrid equation, as a generalization of standard $P(E)$-theory, constitutes the main result of this work.

Now we turn to the details. First, the time-dependent Cooper pair tunneling rate, second order in $E_J$ resp.\ $I_C$, is given by the time integral 
\begin{align}
&\Gamma(t,Q) =  \frac{I_C^2}{8\,e^2}
\int_0^{\infty} dt' \,e^{ - \left(\frac{t'}{\tau_0}\right)^2 }
\nonumber\\[-0.3cm]
\label{Gamma1} \\[-0.2cm]
& \times
\cos\Big{(}\varphi_{\rm cl}(t) - \varphi_{\rm cl}(t-t') - \dot\varphi_{\rm cl}(t)t'  + \frac{2\,e\,(Q - e)}{\hbar \,C} \,t^{'}\Big{)},
\nonumber
\end{align}
where $\tau_0$ is the relaxation time of the environment, i.e.,  
\begin{equation}
\begin{split}
\frac{1}{\tau_0} &\approx\frac{g^{3/4}}{\sqrt{\delta + \sqrt{g}}} 
\sqrt{ 1 + \left|\ln\left(1+\frac{2\sqrt{g}}{\delta}\right) \right| } \frac{2E_C}{\pi\hbar}
\\ &
+ \left( \frac{\pi^2}{3}\frac{g^2}{\delta + g} \frac{k_BT}{E_C} \right)^{1/3} \frac{2E_C}{\pi\hbar}\, .
\end{split}
\label{tau0}
\end{equation}
Here, the dimensionless conductance of the junction is 
\begin{eqnarray}
g = \frac{4R_Q}{R}
\end{eqnarray}
with the parameter $\delta$ being proportional to the inductance
\begin{eqnarray}
\delta = \frac{4\,L}{R^2\,C}\, .
\label{delta}
\end{eqnarray}
In the argument of the $\cos$-function appears the time-dependent phase $\varphi_{\rm cl}(t)$ as  the solution of the classical equation (\ref{RSJ}) with $I_C=0$ and $\xi(t)=0$,
\begin{eqnarray}
\varphi_{\rm cl}(t) = \frac{2\,e\,I_{\rm dc}R}{\hbar} t + \varphi_{\rm ac}\sin(\omega_0t - \theta_0),
\label{phi_cl}
\end{eqnarray}
where
\begin{equation}
    \varphi_{\rm ac} = \frac{2eI_{\rm ac}}{\hbar\omega_0}\sqrt{\frac{R^2+\omega_0^2L^2}{(1-\omega_0^2LC)^2 + \omega_0^2R^2C^2}}, 
\label{phi_ac}
\end{equation}
and
\begin{equation}
\theta_0 = \arctan\frac{\omega_0RC}{1-\omega_0^2LC} - \arctan\frac{\omega_0 L}{R}.    
\nonumber
\end{equation}
Second, the time-dependent effective current $I^*(t)$ appearing in Eq. (\ref{kin}) differs from the input current (\ref{bias}) due to the resonant nature of the circuit of Fig. \ref{Fig1}
at large inductance and has the form
\begin{eqnarray}
I^*(t) = I_{\rm dc} + I^*_{\rm ac}\cos(\omega_0 t - \theta_0^*),
\label{I*}
\end{eqnarray}
with effective amplitude 
\begin{equation}
  I^*_{\rm ac} = \sqrt{\frac{(1-\omega_0^2LC)^2 + \omega_0^2\left( RC + \frac{L}{R} \right)^2}{(1-\omega_0^2LC)^2 + \omega_0^2R^2C^2 }}\,I_{\rm ac},
\label{Iac1}  
\end{equation}
and effective phase shift
\begin{equation}
    \theta_0^* = \arctan\frac{\omega_0\left( RC + \frac{L}{R} \right)}{1-\omega_0^2LC} - \arctan\frac{\omega_0RC}{1-\omega_0^2LC}.
\end{equation}
This current $I^*(t)$ can be roughly interpreted as the effective current seen by the junction. 
The effective temperature $T^*$, also appearing in Eq. (\ref{kin}), is given by the integral
\begin{eqnarray}
T^* = \frac{RC}{k_B} \int\frac{d\omega}{2\pi}\frac{\hbar\omega\coth\frac{\hbar\omega}{2k_BT}}{(1-\omega^2 LC)^2 + \omega^2R^2C^2},
\label{T*}
\end{eqnarray}
which can be approximately written as
\begin{eqnarray}
T^* \approx \left\{ 
\begin{array}{ll}
T + \dfrac{g\,E_C}{\pi^2k_B\sqrt{1-\delta}}\ln\dfrac{1+\sqrt{1-\delta}}{1-\sqrt{1-\delta}}, & \delta < 1,
\nonumber\\[0.5cm]
T + \dfrac{2\,g\,E_C}{\pi^2k_B\sqrt{\delta-1}}\arctan\left(\sqrt{\delta-1}\right), & \delta > 1.
\end{array}
\right.
\label{T**}
\end{eqnarray}
This choice of effective temperature ensures that in the absence of Cooper pair tunneling,
i.e. at $$\Gamma(t,Q)=0,$$ Eq. (\ref{kin}) reduces to the Fokker-Planck equation of the Ornstein-Uhlenbeck process (see e.g. \cite[\S~3.8.4]{GarC2009})  describing charge distribution in the linear circuit consisting of a capacitor $C$, an inductor $L$ and a resistor $R$. In particular, the equilibrium value for vanishing $I^{*}$ of the variance of the charge is given by
\begin{eqnarray}
\left\langle \Big{(}Q-\langle Q \rangle\Big{)}^{2}\right\rangle = C\,k_B\,T^{*}.
\end{eqnarray}
Further details of the derivation of Eq. (\ref{kin}) are presented in Appendix \ref{sec_Gamma}.

\subsection{Range of validity}

Next, let us further elaborate on the range of validity of the kinetic equation (\ref{kin}).

Equation (\ref{kin}) is definitely valid if a time scale separation applies between the sequential Cooper pair transfer and the equilibration of the reservoir, i.e.,   
$$\Gamma(t,Q)\,\tau_0\lesssim 1,$$ \, 
known as the regime of Coulomb blockade.
At $Q=e$, where the rate $\Gamma$ reaches its maximum (see Eq. (\ref{Gamma_ad}) below), this condition translates to 
$$\sqrt{\pi} \,\frac{I_C^2\,\tau_0^2}{16\, e^2} \lesssim 1,$$ 
which leads to a rather strict constraint on the Josephson energy of the junction, namely, 
\begin{eqnarray}
E_J  \lesssim  \frac{\hbar}{\tau_0}.
\label{constraint1}
\end{eqnarray}

The threshold bias current at which the Coulomb blockade is lifted in an isolated Josephson junction at zero temperature is given by $I_{\rm th} = e/RC$. For currents exceeding $I_{\rm th}$, the system experiences Zener tunneling events at the degeneracy charge value $Q=e$ which, however, depend only very weakly on the value of the dissipative charge transfer rate.
Accordingly, one can replace the condition (\ref{constraint1}) with the weaker one, 
$$\Gamma(t,I_{\rm dc}RC)\,\tau_0\lesssim 1 ,$$ 
which further leads to 
\begin{eqnarray}
E_J  \lesssim  \frac{\hbar}{\tau_0} \exp\left( \frac{4\pi^2}{g^2} \frac{\tau_0^2(I_{\rm dc}-I_{\rm th})^2}{e^2}\right)\, .
\label{cond3}
\end{eqnarray}
This condition implies that Eq. (\ref{kin}) is valid if 
$$\frac{g\,e}{2\,\pi\,\tau_0} \lesssim I_{\rm th}$$ 
and for sufficiently small critical currents
\begin{eqnarray}
E_J \lesssim 2E_C\, .
\label{EJEC}
\end{eqnarray}
With the help of  (\ref{tau0}), the first of these conditions reads equivalently
\begin{eqnarray}
\frac{g^{3/4}}{2\pi\sqrt{\delta + \sqrt{g}}} 
+ \left( \frac{1}{24\pi}\frac{g^2}{\delta + g} \frac{k_BT}{E_C} \right)^{1/3} \lesssim 1\, .
\label{condition}
\end{eqnarray}

Condition (\ref{EJEC}) has an immediate consequence: When it is satisfied the potential barrier between neighboring wells of the Josephson cosine potential,  $2E_J$, is lower than the level splitting between localized levels at the bottom of the potential well 
$\sqrt{8\,E_J\,E_C}$. 
Hence, localized levels are absent and
the  wave function in the ground state is strongly delocalized. As we will show below, one can stretch the limits of applicability and
use Eq. (\ref{kin}) for a qualitative analysis of I-V curves even to the range
 $E_J \approx 5\,E_C$. For even higher Josephson energies the half-bandwidth
(\ref{delta0}) becomes too small and the dual Shapiro steps vanish.

Another formal limitation for Eq. (\ref{kin}) is the smallness of the inductance, 
which can be expressed in the form $\delta\ll 1$.
Indeed, the classical dynamics underlining Eq. (\ref{kin}) is the simple relaxation of the charge
$$\dot{Q}(t) + \frac{Q(t)}{R\,C} = I(t) + \xi(t),$$ 
where 
$$Q(t)=\frac{C\,\hbar\,\dot{\varphi}(t)}{2\,e}.$$
It is a good approximation for Eq. (\ref{RSJ}) if 
$$ \frac{R}{L} \gg \omega_0, \frac{1}{RC}. $$
In this case, the inductance just provides the high frequency cutoff for the environment.
To be able to describe resonant systems with $\delta > 1$,
in Eq. (\ref{kin}) we replace the bias current (\ref{bias}) by the effective
current (\ref{I*}) with resonantly enhanced AC amplitude. More details on that are provided in Appendix \ref{sec_Gamma}. 
With such modifications, Eq. (\ref{kin}) can even be used for the analysis of strongly resonant systems with $\delta\gg 1$.
An example of such system is the experimental setup of Ref. \cite{PS2}, in which  $\delta = 169$.

In case of purely Ohmic dissipation, $L =0$,  which is  usually considered in the literature, 
and at zero temperature the condition (\ref{condition}) reduces to $g\lesssim 2\pi$. 
This estimate is consistent
with the rigorous theory predicting a superconductor - insulator phase transition at $g=4$ \cite{Schmid}.
According to that theory, for $g>4$ the dual Shapiro steps are washed out by quantum fluctuations  even in the absence of Zener tunneling \cite{GZ,Hekking,Glazman}. Large inductances (large $\delta$) extend the
validity of Eq. (\ref{kin}) beyond that threshold. Thus, quantum Shapiro steps may be observed 
even at $g>4$ provided $\delta \gg 1$  and the condition (\ref{condition}) is satisfied. 

\subsection{Numerical simulations}

With the kinetic equation (\ref{kin}) at hand, we can now explore the junction dynamics in various regimes of operation. For this purpose, one considers the following observables:

(i) The voltage drop across the junction is given by an integral over the charge, i.e., 
\begin{eqnarray}
V(t) = \int dQ\, \frac{Q}{C}\,W(t,Q).
\label{V}
\end{eqnarray}

(ii) Multiplying Eq. (\ref{kin}) by $Q/C$ and integrating the result over $Q$, we obtain the identity
reflecting the current conservation, namely, 
\begin{equation}
\begin{split}
I(t) &= C\dfrac{dV}{dt}(t) + \frac{V(t)}{R} 
\\
&+ 2\,e\int dQ \,\Big{(}\Gamma(t,Q) - \Gamma(t,-Q) \Big{)}\, W(t,Q).
\end{split}
\label{charge_cons}
\end{equation}
from which we arrive, upon averaging over the time variable, at the dc-current, i.e.,
\begin{eqnarray}
I_{\rm dc} = \frac{V_{\rm dc}}{R} + \overline{ I_J }.
\label{four_point}
\end{eqnarray}
Here, the first term on the right hand side is the dc current flowing through the resistor and second term is
the time averaged Josephson current,
\begin{equation}
\begin{split}
&\frac{\overline{ I_J }}{2\,e} =
\\ 
&\lim_{t\nearrow \infty}\frac{1}{t}
\int_{0}^{t}\mathrm{d}s\int dQ\, \Big{(}\Gamma(s,Q) - \Gamma(s,-Q)\Big{)} W(s,Q).
\end{split}
\label{IJ}
\end{equation}

Now, a numerical analysis necessitates an efficient way to treat the evolution equation (\ref{kin}) for the time-dependent charge distribution $W(Q,t)$. A very convenient method is to convert this equation into an equivalent 
 stochastic differential equation for the charge $Q(t)$, 
constructed in such a way that the charge distribution $W(t,Q)$ fulfills Eq. \ref{kin}. It turns out that this equation takes the form
\begin{equation}
\begin{split}
dQ(t)&+\left(\frac{Q(t)}{R_SC}-I^*(t)\right)\,dt 
\\
&= 2\,e\, \Big{(}dN^+(t) -dN^{-}(t)\Big{)}
\end{split}
\label{eq:stochastic_equation_adiabatic}
\end{equation}
where $dN^\pm$ denote the increments of counting processes fully characterized by the conditions
\begin{equation}
\begin{split}
&dN^{\pm}{}^2(t)=dN^{\pm}(t)
\\
&dN^{+}(t)dN^{-}(t)=0
\\
& \operatorname{E}\big{(}dN^\pm(t)|Q(t)\big{)} = \Gamma (t,\mp Q(t)) dt.
\end{split}    
    \label{eq:expectation_val_n_adiabatic}
\end{equation}
Here, $\operatorname{E}$ denotes the conditional expectation with respect to the value of the charge process at time $t$.
We emphasize that \eqref{eq:stochastic_equation_adiabatic}, \eqref{eq:expectation_val_n_adiabatic} are well defined only if the tunneling rate $\Gamma$ is positive. This is not guaranteed \emph{per se} though. In fact, the rates (\ref{Gamma1}) may oscillate in time and sometimes may even become negative. As we will show below, this can be avoided by imposing further approximations which are in line with the physics we consider in the sequel. Further, in order 
to simplify the simulations and to stress the emergence of the dual and the classical steps, 
we formally put the effective temperature of the resistor in Eq. (\ref{kin}) equal to zero, $T^*=0$.

This way, solving the stochastic differential equation for many charge realizations and performing ensemble and time averages, we obtain 
the desired observables. 
More details about the numerical simulations are given in Appendix~\ref{sec_details_simulation}.

\section{Results and discussion}
\label{results}

\subsection{Adiabatic regime, dual Shapiro steps}

In this section we consider the adiabatic regime, in which the
frequency of the microwave signal and its amplitude are sufficiently small
and the condition
\begin{eqnarray}
\varphi_{\rm ac} (\omega_0\tau_0)^2 \ll 1
\label{cond1}
\end{eqnarray}
is satisfied. In this case, the classical phase changes very little during the relaxation time $\tau_0$ of the environment. This circumstance allows us to approximate
\begin{eqnarray}
\varphi_{\rm cl}(t) - \varphi_{\rm cl}(t-t') - \dot\varphi_{\rm cl}(t)\,t' \approx 0
\label{ad} 
\end{eqnarray}
in the expression for the Cooper pair tunneling rate (\ref{Gamma1}). The approximation has the advantage to reduce the rate (\ref{Gamma1}) to the time independent expression
\begin{eqnarray}
\Gamma_{\rm ad}(Q) = \frac{\sqrt{\pi}}{16}\frac{I_C^2\tau_0}{e^2} \exp\left( - \frac{e^2\tau_0^2(Q-e)^2}{\hbar^2 C^2}\right).
\label{Gamma_ad}
\end{eqnarray}
Eq. (\ref{kin}) with a similar time-independent Cooper pair tunneling rate has been 
originally derived and used to describe the formation of the dual Shapiro steps
by Averin and Likharev \cite{AL}. Sharp steps appear on the I-V curve if 
the rate $\Gamma_{\rm ad}(Q)$ tends to 
\begin{eqnarray}
\Gamma_{\delta}(Q) =  \frac{\pi e E_J^2}{8\hbar E_C} \delta( Q - e )
\label{Gamma_delta}
\end{eqnarray}
We derive this approximation by replacing a unit normalized Gaussian distribution in (\ref{Gamma_ad}) 
with a Dirac-$\delta$. The replacement requires
$$\frac{\hbar^2 C^2}{2\,e^2\,\tau_0^2} \ll 4e^2.$$
We can couch this latter condition into the form
\begin{eqnarray}
\frac{\hbar}{\tau_0} \lesssim 4\,\sqrt{2}\, E_C,
\end{eqnarray}
which is essentially equivalent to (\ref{condition}).

The apparent critical current, or the switching current $I_{\rm sw}$,
is determined by the probability of Zener tunneling $P_Z$ between the lowest and the first excited bands of the Hamiltonian (\ref{HJ})
if the quasicharge grows in time linearly, 
$$Q(t)=\dot{Q}(0)\,t.$$ 
One can show \cite{GZ1} that at the degeneracy point 
$Q=e$ and in the absence of the microwave signal the time derivative
of the charge takes the value 
$$\dot{Q}(0)=I_{\rm dc}-I_{\rm th}.$$
Hence, the probability of Zener tunneling can be estimated as
\begin{equation}
\begin{split}
P_Z &= \exp\left( -\int_{-\infty}^{+\infty} dt\,\Gamma_\delta(Q(t)) \right)
\\
&\approx \exp\left( - \frac{I_Z}{I_{\rm dc} - I_{\rm th}}  \right),
\end{split}
\label{PZ}
\end{equation}
where
\begin{eqnarray}
I_Z  = \frac{\pi E_J}{16 E_C} I_C
\end{eqnarray}
is the Zener current. The switching current in the absence of microwaves
can be estimated in the same way as in Ref. \cite{GZ1}, i.e. by equating
the rate of Zener tunneling from the lowest Bloch band to the higher bands, 
$$\Gamma_\uparrow(I_{\rm dc}) = \frac{I_{\rm dc}}{2\,e}\exp\left(-\frac{I_Z}{I_{\rm dc}-I_{\rm th}}\right),$$
and the rate of Cooper pair tunneling at the equilibrium value of the charge, 
$$\Gamma_\downarrow(I_{\rm dc}) \approx \Gamma_{\rm ad}(I_{\rm dc}\,R\,C).$$
The condition 
$$\Gamma_\uparrow(I_{\rm sw}) = \Gamma_\downarrow(I_{\rm sw})$$ 
leads to
\begin{eqnarray}
I_{\rm sw} \approx \frac{e}{RC} + \left( \frac{\hbar^2 I_Z}{e^2R^2\tau_0^2} \right)^{1/3}.
\label{Isw}
\end{eqnarray}

\begin{figure}
    \centering
      \includegraphics[width=\linewidth]{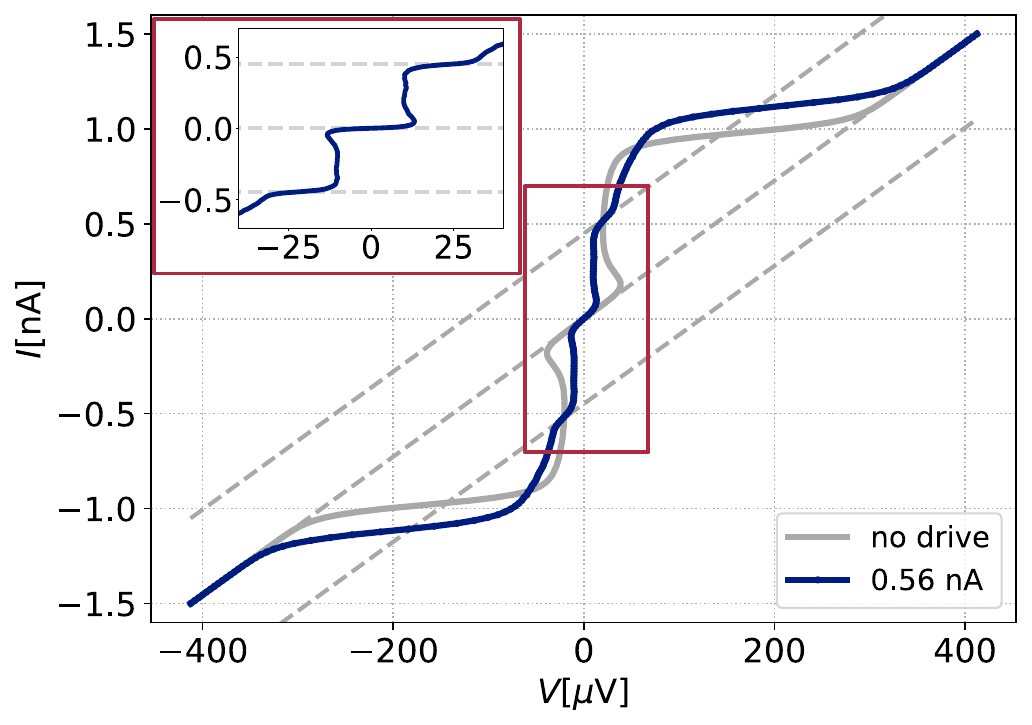}
    \caption{ Comparison of the IV curve of a junction driven with an ac current of $I^*_{\rm ac}=0.56 \rm{nA}$, corresponding to $I_{\rm ac}= 0.56 \rm{nA} $ and a frequency $f_0 = 1.4 \rm{GHz}$ with the undriven IV curve of the same sample. The remaining parameters are $E_J = 60 \mu \rm{eV}$, $E_C = 80 \rm{\mu eV}$, $R=275 \rm{k\Omega}$ $T = 20 \rm{mK}$ and $L= 0.1 \rm{\mu H}$ leading to $\varphi_{\rm ac} (\omega_0\tau_0)^2=0.31$. The inset shows the driven IV curve where the ohmic part was subtracted to increase the visibility of the steps and the gray dashed lines indicate the expected positions of the inverse Shapiro steps. The simulation was performed in the adiabatic limit.}
    \label{fig:l-1nA-1,4GHz}
\end{figure}

\begin{figure}
    \centering
      \includegraphics[width=\linewidth]{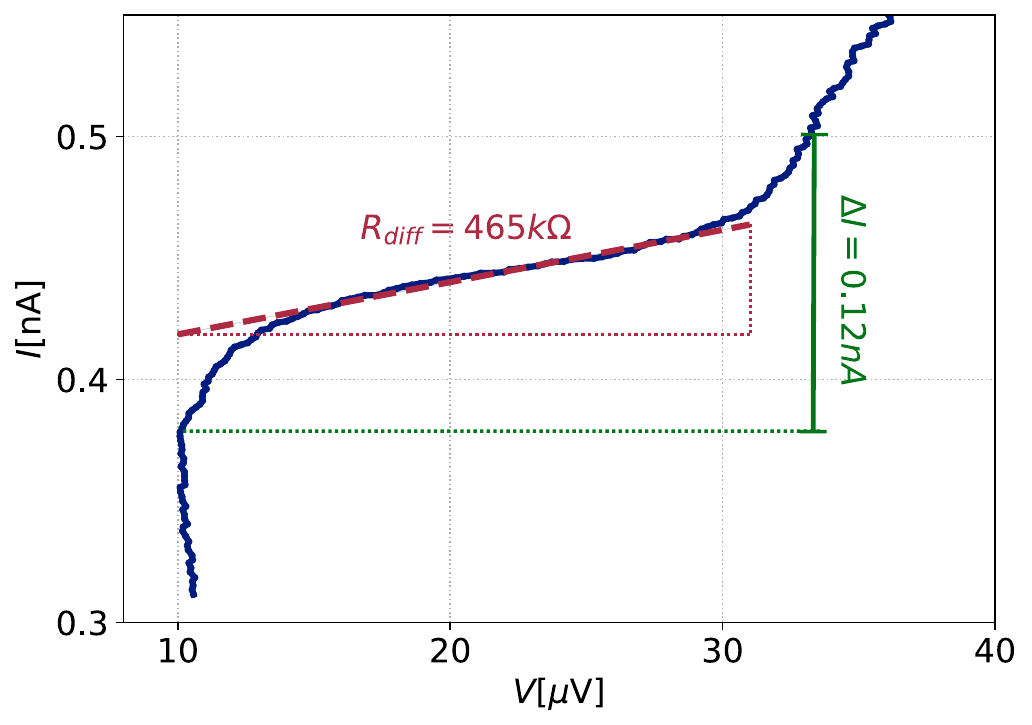}
    \caption{ First inverse step of the IV curve of a junction described by the parameters given in Fig.\ref{fig:l-1nA-1,4GHz}. The width of the step is estimated to be $\Delta I = 0.12 \rm{nA}$ and the differential resistance is $R_{diff} = 465 \rm k \Omega$ The simulation was performed in the adiabatic limit.}
    \label{fig:step_zoom}
\end{figure}

In Fig. \ref{fig:l-1nA-1,4GHz} we plot numerically simulated I-V curves for a junction with
the parameters similar to those of the experiment \cite{PS3}. For the chosen parameters we estimate the environment
relaxation time (\ref{tau0}) as $\tau_0 \approx 1.4\times 10^{-11}$ s
and the phase oscillation amplitude (\ref{phi_ac}) as $\varphi_{\rm ac}\approx 20.3$, which results in rather
small value of the adiabaticity parameter $\varphi_{\rm ac} (\omega_0\tau_0)^2 =0.31$. Thus, the
approximation of time independent Cooper pair tunneling rate (\ref{Gamma_ad}) is justified in this case.
We also estimate the parameter
\begin{eqnarray}
\frac{g^{3/4}}{2\pi\sqrt{\delta + \sqrt{g}}} 
+ \left( \frac{1}{24\pi}\frac{g^2}{\delta + g} \frac{k_BT}{E_C} \right)^{1/3} \approx 0.08.
\end{eqnarray}
Hence, the condition (\ref{condition}) is satisfied and our model can be used for this set
of system parameters. 
Due to the high value of the resistance the system operates in the overdamped regime with
$\delta=0.005$. 
The overall shape of the I-V curve resembles the experimental one. For example, the
switching current, at which the I-V curve without microwave drive switches from the superconducting
to the normal branch of the I-V, equals to 1 nA in our simulation and 
roughly equals to 2 nA in the experiment.
The analytical estimate (\ref{Isw}) gives $I_{\rm sw}=1$ nA in agreement with the numerical simulations.
The Coulomb gap is close to 10 $\mu$V in both cases. In the inset of Fig. \ref{fig:l-1nA-1,4GHz}
we plot the Josephson current $\langle I_J\rangle$ versus the voltage drop across the junction,
subtracting the Ohmic contribution $V_{\rm dc}/R$ from the current according to Eq. (\ref{four_point}).
In this way, the horizontal dual steps at currents $\pm 2ef_0$ become clearly visible. 

In Fig, \ref{fig:step_zoom} we plot the first quantum Shapiro step in large scale.
The Ohmic contribution to the current  has been subtracted, i.e. here we assume
four point measurement scheme like in the inset of Fig. \ref{fig:l-1nA-1,4GHz}.
The width of the step is estimated to be $0.12$ nA and the differential resistance
in the middle of it is $R_{\rm diff}\approx 465$ k$\Omega$.

\begin{figure}
    \centering
    \includegraphics[width=\linewidth]{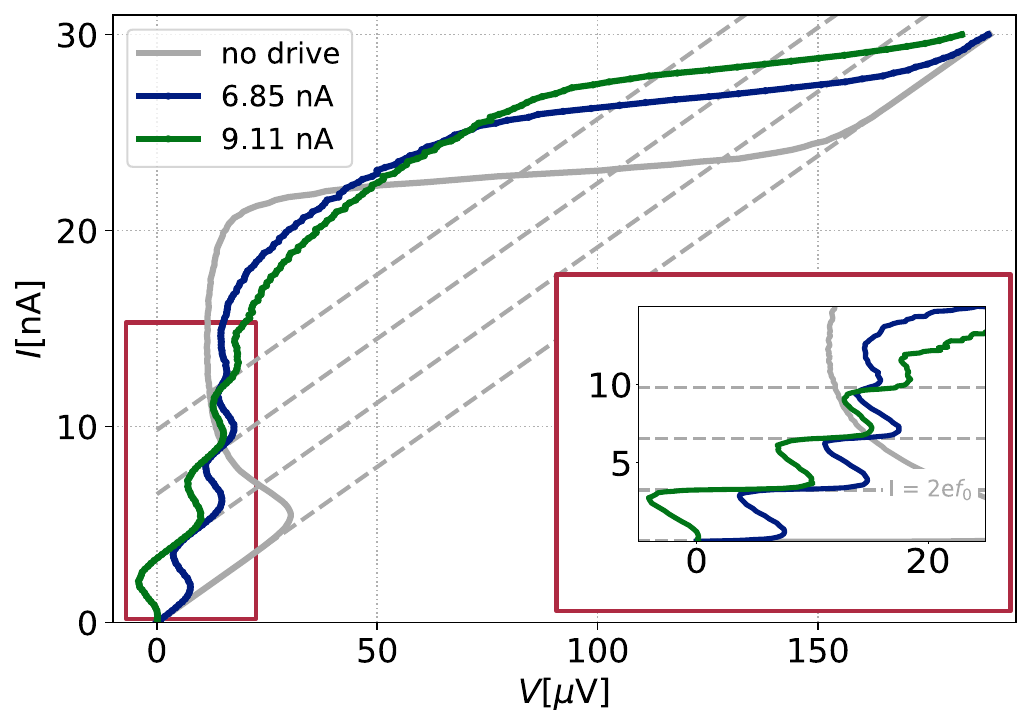}
    \caption{ IV curve of a junction driven with different ac currents ( $I^*_{\rm ac} = 6.85 \rm{nA}$, $I_{\rm ac} = 3.82 \rm{nA} $ and $I^*_{\rm ac} = 9.11 \rm{nA}$, $I_{\rm ac} =5.08 \rm{nA} $) and a frequency $f_0 = 10.215 \rm{GHz}$. The remaining parameters are $E_J = 347 \mu \rm{eV}$, $E_C = 45 \mu \rm{eV}$, $R=6.3 k\Omega$ $T = 200 \rm{mK}$ and $L= 3 \rm{\mu H}$. For comparison also the undriven IV curve is plotted. This leads to  $\varphi_{\rm ac} (\omega_0\tau_0)^2=7$ for $I^*_{\rm ac} = 6.85 \rm{nA}$ and to $\varphi_{\rm ac} (\omega_0\tau_0)^2=9.3$ for $I^*_{\rm ac} = 9.11 \rm{nA}$. The inset shows the IV curve where the ohmic part was substracted to increase the visibility of the steps and the gray dashed lines indicate the expected positions of the inverse Shapiro steps. The simulations were performed in the adiabatic limit.}
    \label{fig:s_10,215_GHz_comparison}
\end{figure}

\begin{figure}
    \centering
    \includegraphics[width=\linewidth]{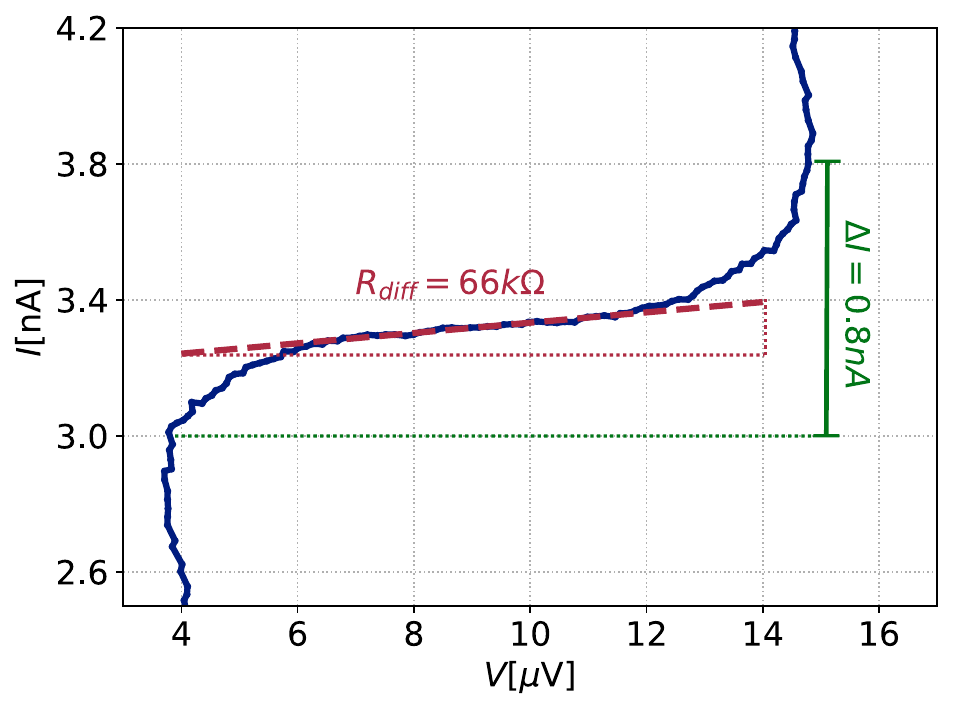}
    \caption{ First inverse Shapiro step of the IV curve of a junction driven with the ac currents  $I^*_{\rm ac} = 6.85 \rm{nA}$, $I_{\rm ac} = 3.82 \rm{nA} $ and a frequency $f_0 = 10.215 \rm{GHz}$. The remaining parameters are given in Fig.\ref{fig:s_10,215_GHz_comparison} where the full IV curve is shown. The with of the step is estimated to be $\Delta I = 0.8 \rm{nA}$ and the differential resistance is $R_{diff} = 66 \rm k\Omega$.}
    \label{fig:step1}
\end{figure}

\begin{figure}
    \centering
    \includegraphics[width=\linewidth]{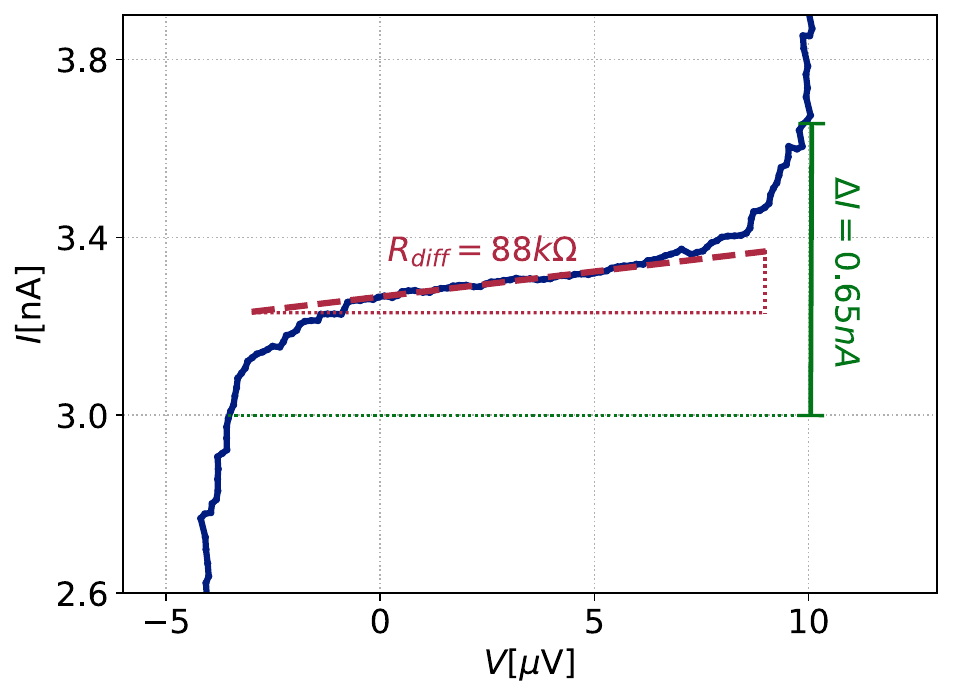}
    \caption{ First inverse Shapiro step of the IV curve of a junction driven with the ac currents $I^*_{\rm ac} = 9.11 \rm{nA}$, $I_{\rm ac} =5.08 \rm{nA}$ and a frequency $f_0 = 10.215 \rm{GHz}$. The remaining parameters are given in Fig.\ref{fig:s_10,215_GHz_comparison} where the full IV curve is shown.  The with of the step is estimated to be $\Delta I = 0.65 \rm{nA}$ and the differential resistance is $R_{diff} = 88 \rm k \Omega$.}
    \label{fig:step2}
\end{figure}
In Fig.\ref{fig:s_10,215_GHz_comparison} we plot another set of the I-V curves
with the parameters taken from the experiment \cite{PS2}. In this case the environment relaxation time (\ref{tau0}) becomes $\tau_0=3.2\times 10^{-11}$ s.
To make the dual steps better visible we apply rather strong microwave currents, 
namely, we choose two values $I_{\rm ac}=3.82$ nA ($I_{\rm ac}^*=6.85$ nA) 
and $I_{\rm ac}=5.08$ nA ($I_{\rm ac}^*=9.11$ nA).
This leads to the AC phase amplitudes $\varphi_{\rm ac}=1.66$ and $\varphi_{\rm ac}=2.2$
and the corresponding values of the adiabaticity parameter
$\varphi_{\rm ac}(\omega_0\tau_0)^2 = 7$ and $\varphi_{\rm ac}(\omega_0\tau_0)^2 = 9.3$.
Thus, the condition (\ref{cond1}) does not hold for this set of parameters.
However, the violation of this condition is not too strong, and we can still hope
to get qualitatively correct results, especially at high bias currents where the term
$$\frac{2\,e\,(Q-e)\,t'}{\hbar\, C} \approx \frac{2\,e\,(I_{\rm dc}RC-e)\,t'}{\hbar\,C}$$ 
in the argument of the cosine
in Eq. (\ref{Gamma1}) becomes large and speeds up the convergence of the integral over time.
The switching current in the absence of microwaves equals to $I_{\rm sw}\approx 23$ nA,
while in the experiment \cite{PS2} a very similar value $I_{\rm sw}\sim 20$ nA has been observed.
The simple analytical expression (\ref{Isw}) predicts a bit higher value $I_{\rm sw}=28$ nA.
The inset again shows the average Josephson current, which is obtained by subtracting the Ohmic
part $V_{\rm dc}/R$ from the main I-V curve. Three sharp steps at the current values
$2ef_0$, $4ef_0$ and $6ef_0$ are clearly visible. The main disagreement between our simulations
and the experiment is in the size of the steps along the voltage axis, namely,
our model significantly overestimates it. In the experiment this size is determined by the critical
voltage is the absence of microwaves, $V_C^{\rm exp}=2.33$ $\mu$eV. In our simulations we
find $V_C\approx 30$ $\mu$V. The reason for this discrepancy is obvious: our model has been developed under
the assumption $E_J\lesssim  2E_C$, while in the experiment one finds $E_J/E_C = 7.7$. Therefore,
one can expect that the size of the steps should be approximately rescaled by a factor $\pi\delta_0/2E_C =  0.02$.
Even though this expected factor is a bit smaller than what we observe in the simulations, 
it gives the correct order of magnitude for rescaling. 
Next, with the parameters given above we estimate
\begin{eqnarray}
\frac{g^{3/4}}{2\pi\sqrt{\delta + \sqrt{g}}} 
+ \left( \frac{1}{24\pi}\frac{g^2}{\delta + g} \frac{k_BT}{E_C} \right)^{1/3} \approx 0.1,
\end{eqnarray}
therefore the condition (\ref{condition}) for the validity of our model is satisfied even though the
resistance $R$ is rather small. This smallness is compensated by a large inductance. 
In contrast to the previous set of parameters, the system of the experiment \cite{PS2}
is strongly resonant with $\delta = 170$. The effective temperature (\ref{T*}) for this sample
is estimated as $T^* = 250$ mK, however as we mentioned above, we put $T^*=0$ in numerical simulations.

In Figs. \ref{fig:step1} and \ref{fig:step2} we present the magnified first dual Shapiro steps for ac currents $I^*_{\rm ac} = 6.85 \rm{nA}$ and $I^*_{\rm ac} = 9.11 \rm{nA}$ . The differential resistances in the middle of the steps are 
$R_{\rm diff, 1}=66$ k$\Omega$ and $R_{\rm diff, 2}=88$ k$\Omega$. 
These values are about 10 times higher than those observed in the experiment \cite{PS2}.
We estimate the width of the steps as $0.8$ nA for the first step and $0.65$ nA for the second step.
In the experiment \cite{PS2} the step width was close to 3 nA.

\subsection{Non-adibatic regime, classcial Shapiro steps}

In this section we demonstrate that Eq. (\ref{kin}) can also describe
classical Shapiro steps in Josephson junctions with a small critical current.
In the non-adibatic or the strong driving limit
\begin{eqnarray}
 (\omega_0\tau_0)^2 \,\varphi_{\rm ac}\gg 1
\label{cond2}
\end{eqnarray}
one can no longer use the approximation (\ref{ad}). Therefore, the Cooper
pair tunneling rate (\ref{Gamma1}) depends on time. In fact, it
quickly oscillates with the typical frequency 
$$\omega_\Gamma\sim \omega_0\,\varphi_{\rm ac}.$$ 
In this regime the system behaves almost classically.
In particular, the charge accumulated in the capacitor stays close to its classical
expression at $I_C=0$,
\begin{equation}
\begin{split}
Q_{\rm cl}(t) &= \frac{\hbar\,C}{2\,e}\dot\varphi_{\rm cl}(t) 
\\
&= I_{\rm dc}RC + Q_{\rm ac}\cos(\omega_0t - \theta_0),
\end{split}
\label{Q_cl}
\end{equation}
where the AC amplitude of charge fluctuations is 
\begin{eqnarray}
Q_{\rm ac} = C \frac{\hbar\omega_0}{2e}\varphi_{\rm ac}.
\end{eqnarray} 
The presence of a small Josephson current causes deviations from the classical behavior \eqref{Q_cl},
resulting in small corrections to the Ohmic I-V curve. These corrections
are the classical Shapiro steps.

We can further simplify the problem if we assume that (\ref{cond2}) and $$\varphi_{\rm ac}\gg 1$$
simultaneously hold. In such a case, the typical frequency of the oscillations of the rate exceeds the driving frequency, $\omega_\Gamma\gg \omega_0$, 
and one can average the rate $\Gamma(t,Q)$ over the time.
Employing the properties of Bessel functions $J_m(x)$,
we can express the exponent containing the classical phases \eqref{phi_cl} as
\begin{equation}
\begin{split}
&e^{i[\varphi_{\rm cl}(t) - \varphi_{\rm cl}(t-t')]} =e^{i\frac{2\, e\,I_{\rm dc}\,R\,t'}{\hbar}}
\\
&\times \sum_{m,n} J_m(\varphi_{\rm ac}) J_n(\varphi_{\rm ac})
e^{i\,m \,(\omega_0\,t - \delta) } e^{-i\,n \,(\omega_0\,(t-t') - \delta) }.
\end{split}
\nonumber
\end{equation}
We observe that only the addends with $m=n$ contribute to the average over $t$ of this sum.
In addition, we can express the term $\dot\varphi_{\rm cl}(t)t'$, which also appears in the argument of the cosine function in Eq. (\ref{Gamma1}), in the form
$$\dot{\varphi}_{\rm cl}(t)\,t' \equiv \frac{2\,e\,Q_{\rm cl}(t)}{C\hbar\omega_0}  t'.$$ 
Upon combining these two approximations we get
\begin{align}
&\lim_{t\nearrow \infty} \frac{1}{t}\int_{0}^{t}\Gamma(t,Q) \Big{|}_{\omega_0\,t=\mathrm{const}} 
\nonumber\\
&=  \sum_{n}\frac{I_C^2}{8\,e^2}J_m(\varphi_{\rm ac})^{2}
\int_0^{\infty} dt' \,e^{ - \left(\frac{t'}{\tau_0}\right)^2 }
\nonumber\\
& \times
\cos\Big{(}\frac{2\, e\,I_{\rm dc}\,R\,t'}{\hbar}+n\,\omega_{0}\,t^{'} + \frac{2\,e\,(Q -Q_{\rm cl}(t)- e)}{\hbar \,C} \,t^{'}\Big{)}.
\nonumber
\end{align}
We are now in the position to evaluate the integral. We conclude that the Cooper pair tunneling rate averaged over the
oscillations with frequencies higher than $\omega_0$ takes the form
\begin{equation}
\begin{split}
&\lim_{t\nearrow \infty} \frac{1}{t}\int_{0}^{t}\Gamma(t,Q) \Big{|}_{\omega_0\,t=\mathrm{const}} 
\\
&=
\tilde{\Gamma}(\varphi_{\rm ac},Q-Q_{\rm ac}\cos(\omega_0t-\theta_0)),
\end{split}
\label{Gamma_hf1}
\end{equation}
where the rate $\tilde\Gamma(\varphi_{\rm ac},Q)$
is expressed via the adiabatic rate (\ref{Gamma_ad}) as
\begin{eqnarray}
\tilde{\Gamma}(Q) = \sum_n J_n^2(\varphi_{\rm ac}) \Gamma_{\rm ad}\left(Q - C\frac{\hbar\omega_0}{2e}n \right).
\label{Gamma_hf}
\end{eqnarray}

The solution of the kinetic equation (\ref{kin}) with the rate (\ref{Gamma_hf1}) takes the form
of a stationary distribution with the center following the classical trajectory,
\begin{eqnarray}
W(t,Q) = \tilde{W}(Q-Q_{\rm ac}\cos(\omega_0t-\theta_0)).
\label{solution_hf}
\end{eqnarray}
The distribution $\tilde {W}(Q)$ can be obtained from
Eq. (\ref{kin}) with the time-dependent current replaced by its dc-component, $I^*(t) \to I_{\rm dc}$,
\begin{eqnarray}
&& \frac{\partial \tilde W}{\partial t}  = 
\frac{\partial}{\partial Q}\left(\frac{Q}{RC} - I_{\rm dc} + \frac{k_BT}{R} \frac{\partial }{\partial Q}\right)\tilde W
\nonumber\\ && 
+\, \tilde\Gamma(Q+2e)\tilde W(Q+2e) + \tilde\Gamma(-Q+2e) \tilde W(Q-2e)
\nonumber\\ &&
-\, \Big{(}\tilde\Gamma(Q) + \tilde\Gamma(-Q)\Big{)} \tilde W(Q),
\label{kin2}
\end{eqnarray}
in the limit $t\to\infty$.
Here we made use of the fact that the classical charge obeys the equation
\begin{eqnarray}
\dot Q_{\rm cl} + \frac{Q_{\rm cl}}{RC} = I^*(t),
\end{eqnarray}
which follows from Eq. (\ref{eq2}) of the Appendix, and this equation, in turn,
leads to the identity
\begin{eqnarray}
&& -\frac{d}{dt}\big(Q_{\rm ac}\cos(\omega_0t - \theta_0)\big) 
\nonumber\\ &&
=\, \frac{Q_{\rm ac}\cos(\omega_0t - \theta_0)}{RC}
-I^*_{\rm ac}\cos(\omega_0t-\theta_0^*).
\nonumber
\end{eqnarray}
The stationary solution of Eq. (\ref{kin2})
describes the formation of classical Shapiro steps at the voltages (\ref{steps}).
The positions of these steps are encoded in the rate $\tilde\Gamma(Q)$ (\ref{Gamma_hf})
and their shape resembles that of the I-V curve of the junction in the absence of
microwave signal. If this I-V curve has a clear vertical superconducting branch then normal classical
Shapiro will appear at voltages (\ref{steps}). In junctions with strong Coulomb blockade
the supercurrent branch is absent, and the classical steps will look like replicas of
the I-V curve with the Coulomb gap.

Substituting the solution (\ref{solution_hf}) in the expression for the voltage (\ref{V}), we obtain
\begin{eqnarray}
V(t) = V_{\rm dc} + \frac{Q_{\rm ac}}{C}\cos(\omega_0t - \varphi_0),
\end{eqnarray}
where the dc component of the voltage reads
\begin{eqnarray}
V_{\rm dc} = \int dQ  \frac{Q}{C} \tilde W(Q).
\end{eqnarray}
The average Josephson current (\ref{IJ}) in this approximation takes the form
\begin{eqnarray}
I_J(V_{\rm dc},\varphi_{\rm ac}) = 2e \int dQ \,\Big{(}\tilde \Gamma(Q) - \tilde\Gamma(-Q)\Big{)}\tilde W(Q).
\label{IJav}
\end{eqnarray}

To clarify the relation between Eq. (\ref{kin2}) and the well known results on the theory of classical
Shapiro steps, we consider the limiting case of very small Josephson critical current such that $\tilde \Gamma(Q)\ll 1/RC$.
In this limit we can replace the charge distribution
$\tilde W(Q)$ in the right hand side of Eq. (\ref{IJav})
by the solution of Eq. (\ref{kin2}) with $\tilde\Gamma=0$. 
This solution has the form $\tilde W(Q) = W_0(Q-I_{\rm dc}RC)$, where $W_0(Q)$ is a simple Gaussian 
\begin{eqnarray}
W_0(Q) = \frac{\exp\left(-\frac{Q^2}{2\,C\,k_B\,T^*}\right)}{\sqrt{2\,\pi \,Ck_B\,T^*}}.
\label{W0}
\end{eqnarray}
At small $\tilde \Gamma$ we can replace $I_{\rm dc}\to V_{\rm dc}/R$ and write the charge distribution
in the form $\tilde W(Q) = W_0(Q-CV_{\rm dc})$. Then,
with the aid of Eq. (\ref{Gamma_hf}), we
can express the DC Josephson current (\ref{IJav}) in presence of microwaves
via the current in the absence of microwaves $I_J^{(0)}$,
\begin{eqnarray}
I_J(V_{\rm dc},\varphi_{\rm ac})  =  
\sum_{n}  J^2_n(\varphi_{\rm ac}) I_J^{(0)}\left(V_{\rm dc}-\frac{\hbar\omega_0}{2e}n,0\right).
\label{TG}
\end{eqnarray}
It is the Tien–Gordon formula \cite{TG}, which well describes the I-V curves of Josephson junctions with small critical current in the classical regime \cite{Lobb,Kot}. 

To demonstrate further relations between our approach and the established results, we consider
the expression for the Josephson current of the junction with small critical current in the absence of microwaves. 
In this case, the current is expressed in terms of the so-called $P(E)$-function \cite{Devoret,IN}:
\begin{eqnarray}
I_J(V_{\rm dc}) = \frac{\pi eE_J^2}{\hbar} \Big{(}P(2eV_{\rm dc}) - P(-2eV_{\rm dc})\Big{)}.
\label{IJ_PE}
\end{eqnarray}
This function is defined in terms of the correlation function of the quantum phase operators,
\begin{eqnarray}
P(E) = \int\frac{dt}{2\pi\hbar}\, e^{i\frac{Et}{\hbar}} \operatorname{Tr} \left(\rho_{\beta} e^{i\hat\varphi(t)} e^{-i\hat\varphi(0)} \right)
\label{PE_corr}
\end{eqnarray}
where $\rho_{\beta}$ is the equilibrium density matrix of the reservoir at inverse temperature $\beta$ (see \cite[\S~3.2.2]{IN} for details).
Upon contrasting Eqs. (\ref{IJav}) and (\ref{IJ_PE}), we observe that they become equivalent if
\begin{eqnarray}
P(E) = \frac{2\hbar}{\pi E_J^2} \int dQ\, \Gamma_{\rm ad}(Q) W_0\left( Q - \frac{CE}{2e} \right).
\label{PE}
\end{eqnarray}
With the $\delta-$function approximation (\ref{Gamma_delta}) for the rate $\Gamma_{\rm hf}(Q)$, the function 
(\ref{PE}) takes the form 
\begin{eqnarray}
P(E) =  \frac{1}{\sqrt{16\pi E_C k_BT^*}}  \exp\left(-\frac{( E - 4E_C )^2}{16 E_Ck_BT^*}\right),
\end{eqnarray}
which is the correct expression for the function (\ref{PE_corr}) in the high resistance limit $R\ll R_Q$ 
and at sufficiently high temperature \cite{Devoret,IN}.

\begin{figure}
    \centering
    \includegraphics[width=\linewidth]{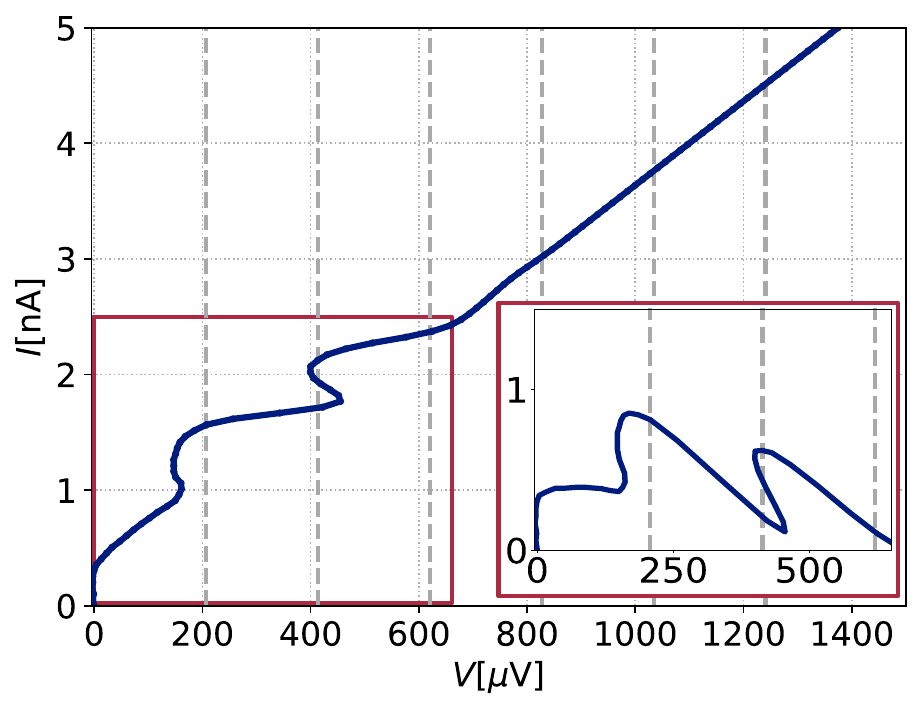}

    \caption{ IV curve of a junction driven with an ac current of $I_{\rm{ac}}=160 \rm{nA}$ and a frequency $f_0 = 100 \rm{GHz}$. The remaining parameters are $E_J = 60 \mu \rm{eV}$, $E_C = 80 \mu \rm{eV}$, $R=275 \rm{k\Omega}$, $T = 20 \rm{mK}$ and $L= 0.1 \mu \rm{H}$ and correspond to those given in Fig.\ref{fig:l-1nA-1,4GHz}. This leads to  $ (\omega_0\tau_0)^2\,\varphi_{\rm ac}=97.2$. Gray dashed lines indicate the expected positions of classical Shapiro steps. The simulation was performed in the classical limit}
    \label{fig:l_160nA_100GHz}
\end{figure}

To investigate the formation of the classical Shapiro steps, 
we have simulated the I-V curves using Eq. (\ref{kin2}) as a starting point. 
In Fig. \ref{fig:l_160nA_100GHz} we show the I-V curve for a junction with the parameters
close to those of the experiment \cite{PS3}, which are given in the previous section.
To achieve the strongly non-adiabatic limit (\ref{cond2}), we have applied a very strong
microwave amplitude $I_{\rm ac}=160$ nA and chosen the frequency $f_0=100$ GHz.
With these parameters we obtain 
$\varphi_{\rm ac} = 1.23$ and $\varphi_{\rm ac}(\omega_0\tau_0)^2 = 97.2$.
Frequencies this high have been used, for example, in the experiment \cite{Kot}. 
In accordance with the predictions of the Tien–Gordon formula (\ref{TG}), we observe the
formation of the classical steps, which resemble the I-V curve without microwave drive 
(grey line in Fig. \ref{fig:l-1nA-1,4GHz}) in shape. They are separated by the voltage intervals
$\hbar\omega_0/2e$ and have varying magnitudes. 
For the chosen parameters the classical steps appear at rather high voltages, which are
equal to and higher than the superconducting gap of aluminum electrodes. For this reason and
because of the heating effects the classical steps may be difficult to observe in the sample
used in the experiment \cite{PS3}.
\begin{figure}
    \centering
    \includegraphics[width=\linewidth]{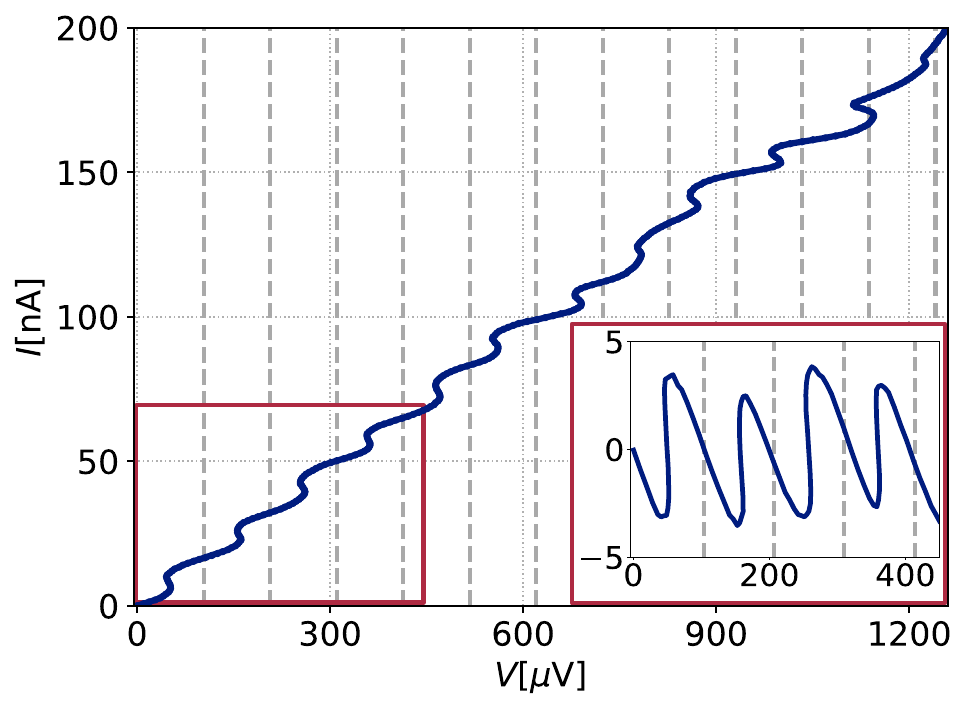}

    \caption{ IV curve of a junction driven with an ac current of $I_{ac}=800 \rm{nA}$ and a frequency $f_0 = 50 \rm{GHz}$. The remaining parameters are $E_J = 347 \mu \rm{eV}$, $E_C = 45 \mu \rm{eV}$, $R=6.3 \rm{k\Omega}$ $T = 200 \rm{mK}$ and $L= 3 \rm{\mu H}$  and correspond to those given in Fig.\ref{fig:s_10,215_GHz_comparison}. This leads to  $(\omega_0\tau_0)^2\,\varphi_{\rm ac} =1402$. Gray dashed lines indicate the expected positions of classical Shapiro steps. The simulation was performed in the classical limit.}
    \label{fig:s_50GHZ_30nA}
\end{figure}
In Fig. \ref{fig:s_50GHZ_30nA} we repeat the same simulations for a junction with the parameters
of the experiment \cite{PS2} and with the strong AC amplitude $I_{\rm ac}=800$ nA and
high frequency $f_0=50$ GHz. This results in
$\varphi_{\rm ac} = 13.7$ and $\varphi_{\rm ac}(\omega_0\tau_0)^2 = 1402$.
In this case, the classical steps are also very similar in shape
to the I-V curve in the absence of microwaves (grey line in Fig. \ref{fig:s_10,215_GHz_comparison}), and
they clearly follow the pattern expected from the Tien–Gordon formula (\ref{TG}).

\section{Conclusion}

In conclusion, we have developed a model of a Josephson junction with small critical current, which describes
both classical and dual (quantum) Shapiro steps within the same formalism. The results of our numerical
simulations agree with the recent experimental data reasonably well. We show that one can potentially
observe both types of Shapiro steps in the same sample by varying the bias current and the amplitude and
the frequency of the microwave signal. However, strong heating of the bias resistors by high bias current
or by high microwave power may prevent such opportunity. Our model accounts for the effect of a large inductor in
the bias circuit, which has been used in recent experiments to limit the high frequency noise of the environment.
Within our model, the effect of this inductor and of the rest of the environment is absorbed into a single parameter -- the relaxation time of the environment $\tau_0$ (\ref{tau0}). 
We show that the dual Shapiro steps are formed at the I-V curve at sufficiently low bias
current, frequency and power of the microwave signal, while the classical steps are
formed in the opposite limit -- sufficiently high bias current, frequency and microwave power.
The crossover between the two regimes is determined by the time $\tau_0$, see the conditions
(\ref{cond1}), (\ref{cond2}) and the expression for the switching current (\ref{Isw}).

A point that deserves to be further explored in future work is the following. 
We numerically integrated our model by averaging over the realizations of the stochastic differential equation \eqref{eq:stochastic_equation_adiabatic}. This is possible as long as \eqref{Gamma1} remains positive at all times thus ensuring that \eqref{kin} is a well-posed classical jump-diffusion equation. In fact, \eqref{Gamma1} may oscillate and temporarily take negative values. In the present paper, we overcome this difficulty by averaging over oscillations. It is, however, well known that quantum master equations beyond the weak coupling scaling limit may actually include non positive rates \cite{HaCrLiAn2014}. The circumstance may describe actual physical phenomena while at the same time restricting the range of physically admissible solutions of a master equation.  Avenues to explore this issue may be provided by adapting the martingale method proposed in \cite{DoMG2022} or, alternatively, directly integrating \eqref{kin} for instance by means of the Quantum Monte Carlo method recently proposed in \cite{ShLi2025}. 

In conclusion, we believe that our model is an important step towards quantitative description of the dual Shapiro steps
and of related quantum phenomena in small area Josephson junctions.

\section{Acknowledgment}
The work of PMG is partially supported by the Finnish CoE in Randomness and Structures (FiRST) of the Research Council of Finland (funding decision number: 346305). MR and JA acknowledge financial support from the Baden-Württemberg Foundation (project QEDHiNet), from the DFG through AN336/13-1, and the BMBF through QSolid.


\appendix

\section{Derivation of Eq. (\ref{kin})}
\label{sec_Gamma}

We consider the system described by the Hamiltonian (\ref{H}) and assume that at the initial moment of time, $t=0$,
the density matrix of the system is factorized. In this case, the density matrix of the junction, $\rho$, evolves in time as \cite{Grabert,SZ,ZaGo2019}
\begin{align}
 \rho(t,\varphi_1,\varphi_2) = 
\int d\tilde{\varphi}_{1}d\tilde{\varphi}_{2} \, J(t, 0;\varphi_1,\varphi_2,\tilde{\varphi}_{1},\tilde{\varphi}_{2}) \rho_0(\tilde{\varphi}_{1},\tilde{\varphi}_{2}),  
\nonumber
\end{align}
where $\rho_0$ is the initial density matrix at time $t=0$. The time evolution kernel $J$ is amenable to the form of the path integral
\begin{eqnarray}
J &=& \int {\cal D}\varphi^+(t') \int {\cal D}\varphi^-(t')
\exp\bigg{(} \frac{i}{\hbar}\int_{0}^t dt' \mathcal{L}
\nonumber\\ && 
-\, i \frac{\hbar}{4e^2} \int_{0}^t dt_1 dt_2 \varphi^-(t_1) Z^{-1}(t_1-t_2) \dot\varphi^+(t_2)
\nonumber\\ &&
-\, \frac{\hbar}{8e^2}\int_{0}^t dt_1 dt_2\,\varphi^-(t_1)G(t_1-t_2)\varphi^-(t_2) \bigg{)}.
\label{J0}
\end{eqnarray}
The argument of the exponential consists of
\begin{align}
\mathcal{L}= \frac{\hbar^2 C}{4e^2}\dot\varphi^- \dot\varphi^+ 
-\, 2E_J\sin\varphi^+\sin\frac{\varphi^-}{2}  + \frac{\hbar I}{2e}\varphi^- 
\nonumber
\end{align}
and
\begin{eqnarray}
&& Z^{-1}(t_1-t_2) = \int\frac{d\omega}{2\pi} \frac{e^{-i\omega (t_1-t_2)}}{Z(\omega)} = \frac{e^{-R(t_1-t_2)/L}}{L} , 
\nonumber\\ &&
G(t_1-t_2) = \int\frac{d\omega}{2\pi}\,{\rm Re}\left[ \frac{\omega}{Z(\omega)} \right] \coth\frac{\hbar\omega}{2k_BT} \, e^{-i\omega (t_1-t_2)},
\nonumber
\end{eqnarray}
Finally,
$$Z(\omega) = R-i\omega L$$ is the impedance of the environment, 
and the "classical" phase $\varphi^+(t')$ and the "quantum" phase $\varphi^-(t')$, 
over which the path integral (\ref{J0}) is taken, satisfy the boundary conditions
\begin{align}
&\varphi^+(t) = \frac{\varphi_1+\varphi_2}{2}, && \varphi^+(0) = \frac{\tilde{\varphi}_{1}+\tilde{\varphi}_{2}}{2}\equiv \varphi_{i}^+,
\nonumber\\
&\varphi^-(t) = \varphi_1 - \varphi_2, && \varphi^-(0) = \tilde{\varphi}_{1} - \tilde{\varphi}_{2} \equiv \varphi_{i}^-.
\nonumber
\end{align} 

\subsection{ the limit $L\to 0$ and $k_BT\gg \hbar/RC$}

In this case, we make the approximations
\begin{equation}
\begin{split}
&Z^{-1}(t_1-t_2) =  \frac{\delta(t_1-t_2-\eta)}{R}, 
\\ 
&G(t_1-t_2) = \frac{2k_BT}{\hbar R} \delta(t_1-t_2),
\end{split}
\label{app1}
\end{equation}
and the action in the path integral (\ref{J0}) becomes local in time. 
Here $\eta>0$ is an infinitesimal positive time shift which ensures that $Z^{-1}(t_1-t_2)\equiv 0$
for $t_1<t_2$ and the causality condition is satisfied. 
It is important to note that the limit of vanishing correlation must be interpreted in the post-point prescription
\begin{equation}
\begin{split}
&    \int_{0}^t dt_1 dt_2 \varphi^-(t_1) Z^{-1}(t_1-t_2) \dot\varphi^+(t_2) 
    \\
&\hspace{0.5cm}    \to\int_{0}^t dt_1\,\varphi^-(t_1^{+}) \dot\varphi^+(t_1)
    \end{split}
    \nonumber
\end{equation}
This is because according to the general properties of response functions, 
\begin{align}
 Z^{-1}(t_1-t_2)\,\,\begin{cases}
 >0 & t_1>t_2
 \\
 = & t_1 \le t_2
 \end{cases}
 \nonumber
\end{align}
to obey causality \cite[\S~2.1]{ZaGo2019}
In this limit, the density matrix satisfies
a first order in time, ``Markovian'', partial differential equation. Namely, by taking the time derivative of Eq. (\ref{J0})
we obtain
\begin{eqnarray}
i\hbar\frac{\partial\rho}{\partial t} &=& 
- 8E_C  \frac{\partial^2 \rho}{\partial\varphi^+ \partial\varphi^-} - i\frac{\hbar}{RC} \varphi^- \frac{\partial \rho}{\partial\varphi^-} 
- \frac{\hbar I(t)}{2e}\varphi^- \rho  
\nonumber\\ &&
-\, i \frac{\hbar k_BT}{4e^2R} (\varphi^{-})^2 \rho  + \frac{\hbar I_C}{e}\sin\varphi^+ \sin\frac{\varphi^-}{2} \rho.
\label{rho2}
\end{eqnarray}
We solve this equation by means of perturbation theory using as small parameter the critical current $I_C$.
Hence, at order zero in the expansion, we set $I_C=0$ and we are able to find the fundamental solution, which we denote by  $J_T$, of the resulting ``thermal'' high-temperature Markovian equation. Such solution satisfies the following boundary condition at equal times:
$$J_T(t',t';\varphi^+,\varphi^-,\varphi_{i}^+,\varphi_{i}^-)=\delta(\varphi^+-\varphi_{i}^+)\delta(\varphi^- -\varphi_{i}^-).$$
We can still write $J_T$ in the form of the path integral (\ref{J0}) by setting $E_J=0$ and using the time local kernels (\ref{app1}). As a result, the path integral becomes Gaussian, and we can explicitly evaluate it:
\begin{equation}
\begin{split}
 J_T =& \frac{\hbar C}{8\pi e^2 K(t-t')}
e^{  i \left(\frac{Q_{\rm cl}(t)}{2e}\varphi^- - \frac{Q_{\rm cl}(t')}{2e}\varphi^-_i\right) } \,
\\ & \times\,
 e^{ i \frac{\hbar}{4e^2 R} \left(\varphi^+ - \varphi_{\rm cl}(t) - \varphi^+_i + \varphi_{\rm cl}(t') \right)\phi^-(t-t') }
\\ & \times\,
e^{ - \frac{k_BT}{4e^2R} (t-t') \big(\phi^-(t-t')\big)^2  }
\\ &\times\,
e^{  \frac{k_BTC}{4 e^2} (\varphi^- - \varphi^-_i) \phi^-(t-t') -  \frac{k_BTC}{8 e^2} \left((\varphi^-)^2 - (\varphi^-_i)^2\right)  }.
\end{split}
\label{JT}
\end{equation}
Here we have introduced the response function of the environment
\begin{eqnarray}
K(t) = RC \left( 1 - e^{-t/RC} \right),
\label{K0}
\end{eqnarray}
and the time dependent combination of the initial and the final quantum phases $\varphi^-,\varphi^-_i$
\begin{eqnarray}
\phi^-(t-t') = RC\frac{\dot K(t-t')\varphi^- - \varphi^-_i}{K(t-t')}.
\label{phi}
\end{eqnarray}
The classical time-dependent phase $\varphi_{\rm cl}(t)$ and the charge $Q_{\rm cl}(t)$ are defined
in Eqs. (\ref{phi_cl},\ref{Q_cl}). 
We also note that the function $J_T$ depends on the difference $\varphi^+ - \varphi^+_i$, which
reflects the phase translational invariance of the system at $I_C=0$.
The validity of the expression (\ref{JT}) can be checked by inserting it back into Eq. (\ref{rho2}).

We need the thermal kernel (\ref{JT}) as an intermediate result, which
allows us to verify the validity of a more general non-Markovian kernel derived below.

To obtain the high temperature version of Eq. (\ref{kin}), we perform several additional steps.
First, we express Eq. (\ref{rho2}) in the integral form
\begin{eqnarray}
&& \rho(t,\varphi^+,\varphi^-) = \frac{\hbar I_C}{e} \int_{-\infty}^t dt' \int d\varphi^+_i d\varphi^-_i
\nonumber\\ && \times\,
J_T(t,t';\varphi^+,\varphi_i;\varphi^-,\varphi^-_i) 
\sin\varphi^+_i \sin\frac{\varphi^-_i}{2} \rho(t',\varphi^+_i,\varphi^-_i),
\nonumber
\end{eqnarray}
and substitute this expression back into the last term of Eq. (\ref{rho2}). Afterwards, 
Eq. (\ref{rho2}) acquires the form
\begin{eqnarray}
&& i\hbar\frac{\partial\rho}{\partial t} =
- 8E_C  \frac{\partial^2 \rho}{\partial\varphi^+ \partial\varphi^-} - i\frac{\hbar}{RC} \varphi^- \frac{\partial \rho}{\partial\varphi^-} 
- \frac{\hbar I(t)}{2e}\varphi^- \rho  
\nonumber\\ &&
-\, i \frac{\hbar k_BT}{4e^2R} (\varphi^-)^2 \rho  
+ \frac{\hbar^2 I_C^2}{e^2}\sin\varphi^+ \sin\frac{\varphi^-}{2} 
\int_{-\infty}^t dt' \int d\varphi^{\pm}_i 
\nonumber\\ && \times\,
J_T(t,t';\varphi^+ - \varphi^+_i;\varphi^-,\varphi^-_i) 
\sin\varphi^+_i \sin\frac{\varphi^-_i}{2} \rho(t',\varphi^{\pm}_i).
\nonumber
\end{eqnarray}
Now we perform the Markovian approximation which consists in performing the time integral over $t'$ in the last term
under the assumption that the density matrix  $\rho$ evolves in time
as if $I_C$ would be equal to zero during the convergence time of the integral. Re-writing
the resulting equation in terms of the charge distribution 
\begin{equation}
\begin{split}
 &W(t,Q) = 
\\
&\int d\varphi^+ d\varphi^- \frac{e^{-i\frac{Q\varphi^-}{2e}}}{4\pi e}\rho\left(t,\varphi^+ + \frac{\varphi^-}{2},\varphi^+ - \frac{\varphi^-}{2}\right), 
\end{split}
\label{rho_Q}
\end{equation}
we arrive at the equation similar to Eq. (\ref{kin}). This set of transformations corresponds to the standard
Bloch-Redfield approximation known in the theory of open quantum systems.  

\subsection{Low temperature and high inductance limit}

Next, we consider the low temperature and high inductance limit, where the system dynamics
is non-Markovian. In this case, 
only a phenomenological treatment is possible. A reasonable approximation is to keep using
the Markovian kinetic equation (\ref{kin}) because it is convenient for numerical simulations of Shapiro steps.
To capture as much of non-Markovian effects as possible, we renormalize the parameters of this equation.
To derive Eq. (\ref{kin}) with modified parameters, we 
start from the approximate Markovian equation for the density matrix similar to the high temperature Eq. (\ref{rho2}),
\begin{eqnarray}
i\hbar\frac{\partial\rho}{\partial t} &=& 
- 8E_C  \frac{\partial^2 \rho}{\partial\varphi^+ \partial\varphi^-} - i\frac{\hbar}{RC} \varphi^- \frac{\partial \rho}{\partial\varphi^-} 
- \frac{\hbar I^*(t)}{2e}\varphi^- \rho  
\nonumber\\ &&
-\, i \frac{\hbar k_BT^*}{4e^2R} (\varphi^-)^2 \rho  + \frac{\hbar I_C}{e}\sin\varphi^+ \sin\frac{\varphi^-}{2} \rho.
\label{rho22}
\end{eqnarray}
Here the modified bias current has the form
\begin{eqnarray}
I^*(t) = I_{\rm dc} + I^*_{\rm ac}\cos(\omega_0 t - \varphi_0^*),
\end{eqnarray}
where the amplitude $I^*_{\rm ac}$ and the phase $\varphi_0^*$ are defined in Eq. (\ref{Iac1}).
These parameters  are chosen in such a way that the solutions the  equation
\begin{eqnarray}
C \frac{\hbar\ddot\varphi_{\rm cl}}{2e} + \int_{-\infty}^t dt' \frac{e^{-\frac{R(t-t')}{L}}}{L} \frac{\hbar\dot\varphi_{\rm cl}(t')}{2e} = I(t)
\label{eq1}
\end{eqnarray} 
and of the equation
\begin{eqnarray}
C \frac{\hbar\ddot\varphi_{\rm cl}}{2e} + \frac{1}{R} \frac{\hbar\dot\varphi_{\rm cl}}{2e} = I^*(t)
\label{eq2}
\end{eqnarray} 
coincide and equal to the classical phase (\ref{phi_cl}). Thus, although Eq. (\ref{rho22}) 
formally describes the overdamped classical
dynamics of Eq. (\ref{eq2}), it approximately captures the resonant behavior of Eq. (\ref{eq1}) at large inductance $L$
via the enhancement of the AC current amplitude (\ref{Iac1}). At low frequencies $\omega\ll R/L$ the two classical equations (\ref{eq1}) and (\ref{eq2}) 
become equivalent. Therefore, Eq. (\ref{rho22}), based on the overdamped classical equation (\ref{eq2}), should correctly capture the low frequency
quantum dynamics, which is important for obtaining correct DC I-V curve of the junction.

Next, we treat the non-linear term $\propto I_C$ in Eq. (\ref{rho22}) by means of
perturbation theory in the same way as we did above for the Markovian equation (\ref{rho2}).
We find the evolution kernel $J$ in the zeroth order
by solving the Gaussian path integral (\ref{J0}) with $I_C=0$ and with the boundary conditions
\begin{eqnarray}
\varphi^+(t) = \frac{\varphi_1+\varphi_2}{2}, \;\; \varphi^+(t') = \frac{\varphi_{1i}+\varphi_{2i}}{2},
\nonumber\\
\varphi^-(t) = \varphi_1 - \varphi_2,\;\; \varphi^-(t') = \varphi_{1i} - \varphi_{2i}.
\end{eqnarray} 
It is a straightforward, but rather lengthy, procedure, here we follow the recipes of the review \cite{Grabert}. 
To express the result, we
define the response function of the environment at finite inductance, 
\begin{eqnarray}
K(t) = 
\frac{\kappa_+\left( 1 - e^{-\kappa_- t} \right)}{\kappa_-(\kappa_+ - \kappa_-)}
- \frac{\kappa_-\left( 1 - e^{-\kappa_+ t} \right)}{\kappa_+(\kappa_+ - \kappa_-)}.
\label{K}
\end{eqnarray}
Here the two relaxation rates are defined as
\begin{eqnarray}
\kappa_\pm = \frac{R}{2L} \pm \sqrt{\frac{R^2}{4L^2} - \frac{1}{LC}}.
\end{eqnarray}
For large $L> R^2C/2$ ($\delta > 1$) the rates $\kappa_\pm$ become complex and the response function oscillates in time. 
In contrast, for small inductance $L\to 0$ one finds $\kappa_+ \to\infty$, $\kappa_-=1/RC$ and the function $K(t)$ reduces to the form (\ref{K0}).
Evaluating the path integral we obtain the following expression for the function $J$:
\begin{align}
\begin{split}
J =& \frac{\hbar C}{8\pi e^2 K(t-t')}
e^{  i  \left(\frac{Q_{\rm cl}(t)}{2e}\varphi^- - \frac{Q_{\rm cl}(t')}{2e}\varphi^-_i\right) }
\\ & \times\,
e^{ i \frac{\hbar}{4e^2R} \big(\varphi^+ - \varphi_{\rm cl}(t) - \varphi^+_i + \varphi_{\rm cl}(t')\big)\phi^-(t-t')  }
\\ & \times\,
e^{ - f_1(t-t')\left(\phi^-(t-t')\right)^2 }
\\ & \times\,
e^{ - f_2(t-t')(\varphi^-)^2 - f_3(t-t')\varphi^-\phi^-(t-t') }.
\end{split}
\label{J}
\end{align}
Here the function $\phi^-(t-t')$ is still defined by Eq. (\ref{phi}), but
with the modified response function (\ref{K}), and the functions $f_1,f_2,f_3$ are expressed as
\begin{align}
\begin{split}
f_1(t)& = f(t) - K(t)\dot f(t) + \frac{K^2(t)}{2}\ddot f(0),
\\
f_2(t) &= R^2C^2\left( -\dot K(t)\ddot f(t) + \frac{1+\dot K^2(t)}{2} \ddot f(0) \right),
\\
f_3(t) &= R\,C \Big{(} 1-\dot K(t)\Big{)}\,\dot f(t) 
\\
&-\, R\,C \, K(t)\Big{(}\ddot f(t) + \dot K(t) \ddot f(0) \Big{)} . 
\end{split}
\label{f123}
\end{align}
Finally, the function $f(t)$, appearing in the expressions above, is given by the integral
\begin{eqnarray}
f(t) = \frac{\hbar}{4e^2 R} \int\frac{d\omega}{2\pi}\frac{\coth\frac{\hbar\omega}{2k_BT}( 1-\cos\omega t )}{\omega\big{(}(1-\omega^2 LC)^2 + \omega^2R^2C^2\big{)}}.
\end{eqnarray}
A high temperatures $k_BT \gg \hbar/RC$ and for $L=0$ one finds
\begin{eqnarray}
f(t)  &=& \frac{k_BT}{4e^2R}\left(t - RC\left( 1-e^{-\frac{t}{RC}} \right)\right),
\end{eqnarray}
and the fundamental solution (\ref{J}) reduces to the thermal kernel (\ref{JT}).
As we noted above, this fact provides an additional verification of the validity of Eq. (\ref{J})
and helps to resolve the uncertainties in the choice of the saddle point solutions 
while evaluating the path integral.

Next, we repeat the steps taken previously and formally express Eq. (\ref{rho22}) in the integral form
\begin{eqnarray}
&& \rho(t,\varphi^+,\varphi^-) = \frac{\hbar I_C}{e} \int_{-\infty}^t dt' \int d\varphi^+_i d\varphi^-_i
\nonumber\\ && \times\,
J(t,t';\varphi^+,\varphi_i;\varphi^-,\varphi^-_i) 
\sin\varphi^+_i \sin\frac{\varphi^-_i}{2} \rho(t',\varphi^+_i,\varphi^-_i).
\nonumber
\end{eqnarray}
Substituting this integral in the last term of Eq. (\ref{rho22}), we obtain
\begin{eqnarray}
&& i\hbar\frac{\partial\rho}{\partial t} =
- 8E_C  \frac{\partial^2 \rho}{\partial\varphi^+ \partial\varphi^-} - i\frac{\hbar}{RC} \varphi^- \frac{\partial \rho}{\partial\varphi^-} 
- \frac{\hbar I^*(t)}{2e}\varphi^- \rho  
\nonumber\\ &&
-\, i \frac{\hbar k_BT^*}{4e^2R} (\varphi^-)^2 \rho  
+ \frac{\hbar^2 I_C^2}{e^2}\sin\varphi^+ \sin\frac{\varphi^-}{2} 
\int_{-\infty}^t dt' \int d\varphi^\pm_i
\nonumber\\ && \times\,
J(t,t';\varphi^+ - \varphi^+_i;\varphi^-,\varphi^-_i) 
\sin\varphi^+_i \sin\frac{\varphi^-_i}{2} \rho(t',\varphi^\pm_i).
\nonumber\\
\label{rho222}
\end{eqnarray}

Next, we explain in some detail how  Eq. (\ref{kin}) is derived from Eq. (\ref{rho222}).
First, we note that in the limit of small $I_C$, which we are considering here, we can neglect the dependence of $\rho(t',\varphi^+_i,\varphi^-_i)$
in the last term of Eq. (\ref{rho222})
on the phase $\varphi^+_i$ because at $I_C=0$ the translational invariance holds. Next, writing the product of two sines in the form
$$\sin\varphi^+\sin\varphi^+_i = \frac{-\cos(\varphi^+ + \varphi^+_i) + \cos(\varphi^+ - \varphi^+_i)}{2}$$ we observe that the
term containing $\cos(\varphi^+ + \varphi^+_i)$ vanishes after the integration over $\varphi^+_i$ because $J$ depends on the difference
$\varphi^+ - \varphi^+_i$ and $\rho(t',\varphi^+_i,\varphi^-_i)$ does not depend on $\varphi^+_i$.
Thus, Eq. (\ref{rho222}) can be written in the form
\begin{eqnarray}
&& i\hbar\frac{\partial\rho}{\partial t} =
- 8E_C  \frac{\partial^2 \rho}{\partial\varphi^+ \partial\varphi^-} - i\frac{\hbar}{RC} \varphi^- \frac{\partial \rho}{\partial\varphi^-} 
- \frac{\hbar I^*(t)}{2e}\varphi^- \rho  
\nonumber\\ &&
-\, i \frac{\hbar k_BT^*}{4e^2R} (\varphi^-)^2 \rho  
+ \frac{\hbar^2 I_C^2}{2e^2} \sin\frac{\varphi^-}{2} 
\int_{-\infty}^t dt' \int d\varphi^+_i d\varphi^-_i
\nonumber\\ && \times\,
J(t,t';\varphi^+ - \varphi^+_i;\varphi^-,\varphi^-_i) 
\cos(\varphi^+ - \varphi^+_i) \sin\frac{\varphi^-_i}{2} 
\nonumber\\ && \times\,
\rho(t',\varphi^+,\varphi^-_i).
\label{rho3}
\end{eqnarray}

Now we re-write Eq. (\ref{rho3}) in terms of the charge distribution (\ref{rho_Q}), which gives
\begin{eqnarray}
&& \frac{\partial W}{\partial t}(t,Q) =
\frac{\partial}{\partial Q} \left( \frac{Q}{RC} - I^*(t) + \frac{k_BT^*}{R} \frac{\partial }{\partial Q} \right)W(t,Q)
\nonumber\\ && 
+\, \frac{I_C^2}{16e^2}\frac{1}{4\pi e}\sum_{\nu=\pm 1}\sum_{\mu_1,\mu_2=\pm 1} (-1)^{\frac{\mu_1+\mu_2-2}{2}}  
\int_{-\infty}^t dt' \int dQ'
\nonumber\\ &&\times\,
\sqrt{\frac{\pi}{f_2(t-t')}}
e^{-\frac{16e^4R^2}{\hbar^2} \left( f_1(t-t') - \frac{f_3^2(t-t')}{4f_2(t-t')} \right)  }
\nonumber\\ &&\times\,
e^{ i \nu \left( \varphi_{\rm cl}(t) - \varphi_{\rm cl}(t') - \dot\varphi_{\rm cl}(t')  K(t-t') \right) }
\nonumber\\ &&\times\,
e^{ -\frac{ \left( Q - \mu_1 e - Q_{\rm cl}(t) - \left(Q' - Q_{\rm cl}(t') + \mu_2e\right) \dot K(t-t')  \right)^2  }{16e^2f_2(t-t')} }
\nonumber\\ && \times\,
e^{i\nu \frac{eR}{\hbar} \frac{f_3(t-t')}{f_2(t-t')}  \left( Q - Q_{\rm cl}(t) - \mu_1e - (Q' - Q_{\rm cl}(t') + \mu_2e)\dot K(t-t')  \right)  }
\nonumber\\ && \times\,
e^{i \nu\mu_2 \frac{2e^2}{\hbar C}  K(t-t') } e^{i\nu\frac{2e}{\hbar C}K(t-t') Q' } W(t',Q').
\label{rho4}
\end{eqnarray}
Here the indexes $\nu,\mu_1,\mu_2$ come from expressing cosines and sines in the form 
\begin{equation}
\begin{split}
&\cos(\varphi^+-\varphi^+_i) = \sum_{\nu=\pm 1} \frac{e^{i\nu(\varphi^+-\varphi^+_i)}}{2},
\\
&\sin\frac{\varphi^-}{2} = \sum_{\mu_1=\pm 1} (-1)^{(\mu_1-1)/2} \frac{e^{i\mu_1\varphi^-/2}}{2i}, 
\\
&\sin\frac{\varphi^-_i}{2} = \sum_{\mu_2=\pm 1} (-1)^{(\mu_2-1)/2} \frac{e^{i\mu_2\varphi^-_i/2}}{2i}.
\end{split}
\end{equation} 
The next step is to shift the integration parameter in Eq. (\ref{rho4}) as
\begin{eqnarray}
Q' \to Q' - \mu_2e   + Q_{\rm cl}(t') + \frac{ Q - \mu_1 e - Q_{\rm cl}(t) }{\dot K(t-t')}
\nonumber
\end{eqnarray}
and write it in the form
\begin{eqnarray}
&& \frac{\partial W}{\partial t}(t,Q) =
\frac{\partial}{\partial Q} \left( \frac{Q}{RC} - I^*(t) + \frac{k_BT^*}{R} \frac{\partial }{\partial Q} \right)W(t,Q)
\nonumber\\ && 
+\, \frac{I_C^2}{16e^2}\frac{1}{4\pi e}\sum_{\nu=\pm 1}\sum_{\mu_1,\mu_2=\pm 1} (-1)^{\frac{\mu_1+\mu_2-2}{2}}  
\int_{-\infty}^t dt' \int dQ'
\nonumber\\ &&\times\,
\sqrt{\frac{\pi}{f_2(t-t')}}
e^{-\frac{16e^4R^2}{\hbar^2} \left( f_1(t-t') - \frac{f_3^2(t-t')}{4f_2(t-t')} \right)  }
\nonumber\\ &&\times\,
e^{ i \nu \left( \varphi_{\rm cl}(t) - \varphi_{\rm cl}(t') - \dot\varphi_{\rm cl}(t)  \frac{K(t-t')}{\dot K(t-t')} \right) }
e^{i\nu\frac{2e}{\hbar C} \frac{K(t-t')}{\dot K(t-t')} (Q - \mu_1 e)  } 
\nonumber\\ &&\times\,
e^{ -\frac{ \dot K^2(t-t')  }{16e^2f_2(t-t')}{Q'}^2 }
e^{ i\nu \frac{2eR}{\hbar}\left( \frac{K(t-t')}{RC} -   \frac{f_3(t-t')\dot K(t-t')}{2f_2(t-t')}  \right) Q' }
\nonumber\\ && \times\,
W\left(t', Q' - \mu_2e   + Q_{\rm cl}(t') + \frac{ Q - \mu_1 e - Q_{\rm cl}(t) }{\dot K(t-t')} \right).
\nonumber\\
\label{rho5}
\end{eqnarray}

To proceed further, 
we assume that the function $f(t)$ quickly grows with time so that one can make Markovian approximation and
get rid of the time integral in the last term of Eq. (\ref{rho5}). We also assume that the charge distribution weakly changes
during the time scale $\tau_0$, at which the time integral converges. Then at times $t-t' \lesssim \tau_0$ we can approximate
\begin{eqnarray}
&& \sqrt{\frac{\pi}{f_2(t)}} e^{ -\frac{ \dot K^2(t)  }{16e^2f_2(t)}{Q'}^2 }
e^{ i\nu \frac{2eR}{\hbar}\left( \frac{K(t)}{RC} -   \frac{f_3(t)\dot K(t)}{2f_2(t)}  \right) Q' }
\nonumber\\ &&
\approx \frac{ 4\pi e }{|\dot K(t)|} 
e^{ - \frac{16e^4R^2}{\hbar^2} f_2(t) \left( \frac{K(t)}{RC\dot K(t)} -   \frac{f_3(t)}{2f_2(t)}  \right)^2 }
\delta(Q').
\nonumber\\
\label{app}
\end{eqnarray}
This approximation is possible if $8f_2(\tau_0)\lesssim 1$, in which case the spread of the charge $Q'$ during the 
integration time $\tau_0$ is much less than $e$, i.e. $\sqrt{\langle \delta {Q'}^2 \rangle_{\tau_0}}\ll e$.
The condition $8f_2(\tau_0)\lesssim 1$ in combination with the expression for the time $\tau_0$ (\ref{tau0}) again leads to
the condition (\ref{condition}), which ensures the validity of our approximations at all stages.
The approximation (\ref{app}) allows one to carry out the integration over $Q'$ in Eq. (\ref{rho5}), resulting in
\begin{eqnarray}
&& \frac{\partial W}{\partial t}(t,Q) =
\frac{\partial}{\partial Q} \left( \frac{Q}{RC} - I^*(t) + \frac{k_BT^*}{R} \frac{\partial }{\partial Q} \right)W(t,Q)
\nonumber\\ && 
+\, \frac{I_C^2}{16e^2}\sum_{\nu=\pm 1}\sum_{\mu_1,\mu_2=\pm 1} (-1)^{\frac{\mu_1+\mu_2-2}{2}}   
\nonumber\\ &&\times\,
\int_{-\infty}^t dt'
\frac{1}{|\dot K(t-t')|} e^{- F(t-t')} e^{i\nu\frac{2e}{\hbar C}\frac{K(t-t')}{\dot K(t-t')} (Q - \mu_1 e) }
\nonumber\\ &&\times\,
e^{ i \nu \left( \varphi_{\rm cl}(t) - \varphi_{\rm cl}(t') - \dot\varphi_{\rm cl}(t)  \frac{K(t-t')}{\dot K(t-t')} \right) }
\nonumber\\ && \times\,
W\left(t',   \frac{ Q - \mu_1 e - Q_{\rm cl}(t) }{\dot K(t-t')} - \mu_2e   + Q_{\rm cl}(t') \right).
\label{rho6}
\end{eqnarray}
Here the function $F(t)$ is defined as
\begin{equation}
\begin{split}
F(t) &= 
\\
&\frac{64\pi^2}{g^2} 
\left(
f_1(t) + \frac{K^2(t) f_2(t)}{R^2C^2\dot K^2(t)} - \frac{K(t)f_3(t)}{RC\dot K(t)}
\right).
\end{split}
\label{F}
\end{equation}

We assume that the time integral in Eq. (\ref{rho6}) quickly converges. To determine the
corresponding time scale $\tau_0$ we expand the function $F(t)$ at small times.
In the limit of small inductance, such that $\delta \lesssim 2\sqrt{g}$ and $\kappa_+ \gg \kappa_-$, the relevant times 
belong to the interval $\kappa_+^{-1} < t < \kappa_-^{-1}$, in which one can approximate
\begin{equation}
F(t) \approx \frac{4gE_C^2}{\pi^2\hbar^2} t^2\ln(\kappa_+ t) + \frac{8}{3\pi} g \frac{ k_BT E_C^2}{\hbar^3} t^3 .
\label{exp1}
\end{equation}
For higher inductances, $2\sqrt{g} \lesssim \delta \leq 1$, 
the relevant time scales are  $t\lesssim \kappa_+^{-1}$, where one finds
\begin{align}
&F(t) \approx  \frac{4}{\pi^2} \frac{g^2}{\delta} \frac{k_BTE_C^3}{\hbar^4} t^4+\frac{16g^3 E_C^4}{\pi^4\hbar^4\delta^2}
\nonumber\\[-0.3cm]
\label{exp2}
\\[-0.1cm]
&\times\bigg( \ln\frac{\sqrt{LC}}{t} + \frac{7}{4} - \gamma- \frac{2-\delta}{4\sqrt{1-\delta}}\ln\frac{1 + \sqrt{1-\delta}}{1 - \sqrt{1-\delta}} \bigg) t^4.
\nonumber
\end{align}
Combining the two expressions and making some additional approximations, 
we conclude that at short times the function $F(t)$ can be reasonably well approximated  as
\begin{align}
F(t) = \begin{cases}
 \dfrac{t^2}{\tau_0^2}, & \delta < 2\sqrt{g},
\\[0.4cm]
\dfrac{t^4}{\tau_0^4},  & \delta > 2\sqrt{g},
\end{cases}
\label{taylor}
\end{align}
where the relaxation time $\tau_0$ is defined by Eq. (\ref{tau0}) in the main text. Eq.  (\ref{tau0})
has been derived by approximate interpolation between the expansions (\ref{exp1}) and (\ref{exp2}).

Next, we observe that since the condition $\kappa_-\tau_0\lesssim 1$ holds, one can expand the 
function $K(t-t')$ at small time differences $t-t'$, which results in
\begin{eqnarray}
\dot K(t-t') \approx 1,
\;\;\;
\frac{K(t-t')}{\dot K(t-t')} \approx t-t'. 
\end{eqnarray} 
After that, Eq. (\ref{rho6}) acquires the form 
\begin{eqnarray}
&& \frac{\partial W}{\partial t}(t,Q)=
\frac{\partial}{\partial Q} \left(  \frac{Q}{RC} - I^*(t)  + \frac{k_BT^*}{R} \frac{\partial}{\partial Q} \right)W(t,Q)
\nonumber\\ && 
+\, \frac{I_C^2}{16e^2}\sum_{\nu=\pm 1}\sum_{\mu_1,\mu_2=\pm 1} (-1)^{\frac{\mu_1+\mu_2-2}{2}}   
\nonumber\\ &&\times\,
\int_{-\infty}^t dt'\, e^{- F(t-t')} e^{i\nu\frac{2e}{\hbar C} (Q - \mu_1 e)(t-t') }
\nonumber\\ &&\times\,
e^{ i \nu \left( \varphi_{\rm cl}(t) - \varphi_{\rm cl}(t') - \dot\varphi_{\rm cl}(t)(t-t') \right) }
\nonumber\\ && \times\,
W\left(t',   Q - Q_{\rm cl}(t)  + Q_{\rm cl}(t') - (\mu_1 + \mu_2) e \right).
\label{rho7}
\end{eqnarray}

Now we make the following observation:
if we neglect the last term $\propto I_C^2$ in Eq. (\ref{rho7}), then its solution takes a simple form $W(t,Q) = W_0(Q-Q_{\rm cl}(t))$, where
$W_0(Q)$ is the stationary solution defined in Eq. (\ref{W0}), which just follows the classical solution without changing its shape. 
Since we are considering the perturbation theory in $I_C$, we can assume
that during the short time interval between $t$ and $t'$  the distribution $W$, appearing in the last term under the time integral,
evolves in time in the same way as the solution at $I_C=0$, i.e. the function $W_0(Q-Q_{\rm cl}(t))$, does.
Under this assumption we find
\begin{eqnarray}
&& W\left(t', Q - (\mu_1 + \mu_2) e - Q_{\rm cl}(t)  + Q_{\rm cl}(t') \right)
\nonumber\\ &&
=\, W\left(t, Q - (\mu_1 + \mu_2) e \right).
\end{eqnarray}
Performing this replacement together with the summation over the indexes $\nu,\mu_1,\mu_2$  in Eq. (\ref{rho7}), we transform it 
to the form (\ref{kin}) with the Cooper pair tunneling rate taking the form 
\begin{align}
& \Gamma(t,Q) =  \frac{I_C^2}{8e^2}
\int_0^{\infty} dt' e^{ - F(t') }
\nonumber\\[-0.3cm]
\label{Gamma_exact}
\\[-0.2cm]
& \times\,
\cos\Big{(} \varphi_{\rm cl}(t) - \varphi_{\rm cl}(t-t') - \dot\varphi_{\rm cl}(t)t'  + \frac{2e}{\hbar C}(Q - e)t'  \Big{)}.
\nonumber
\end{align}
The integration over time in this formula cannot be carried out exactly because of
the complicated time dependence of the function $F(t)$, which grows exponentially at long times $\kappa_-t \gtrsim 1$. 
Moreover, such integration would make little sense anyway
because we have already made several approximations based on the assumption that the time $t'$ is small.
Therefore, the short time scale $\tau_0$, at which the time integral in Eq. (\ref{Gamma_exact}) converges, 
is the only parameter which can be reliably extracted from the model. 
For this reason, we simply approximate $F(t) \approx t^2/\tau_0^2$ for the whole range of parameters, 
which leads to the expression for the rate (\ref{Gamma1}) given in the main text.

\section{details of the numerical implementation } 
\label{sec_details_simulation}
In order to determine the IV curves, the stochastic differential equation given in  eq. \ref{eq:stochastic_equation_adiabatic} has to be solved numerically for different realizations of the charge evolution.
To do so, at each time step $t$ the probability $P$ of making a jump is determined by
\begin{eqnarray}
P(t) = 1- \exp\left(-\Big{(}\Gamma(t,Q)+\Gamma(t,-Q)\Big{)}dt\right)
\end{eqnarray}
where the time step is chosen sufficiently small to ensure that the probability to jump twice within $dt$ is negligible.
If a jump occurs, we update the value of the charge $Q$ by increasing or decreasing it by the charge of a Cooper pair
\begin{eqnarray}
    Q(t+dt)= Q(t)\pm 2e,
\end{eqnarray}
with probability determined by the ratio $\Gamma(Q)/(\Gamma(Q)+\Gamma(-Q))$.
If no jump takes place, we compute the next value of $Q$ according to the Euler scheme 
\begin{eqnarray}
Q(t+dt) = -\left(\frac{Q(t)}{RC}-I(t)\right)dt.
\end{eqnarray}
this process is repeated for 100 charge ensembles to determine the ensemble average of the evolution of the charge.
Furthermore the time evolution is determined for several oscillations of $\omega_0$ to ensure the average of the Charge is converged.
The Voltage is then given by $V = \langle Q \rangle/C$.
For the simulations in this paper a typical time step of $dt = 0.05/\rm max(\Gamma(Q))$ is used and the evolution of the charge is simulated for between $10^5-10^6$ time steps to ensure convergence. For more information on the numerical implementation see \cite{code}.

\bibliography{Quantum_Shapiro}
\end{document}